\documentclass[11pt]{article}
\usepackage{amssymb}
\usepackage{citesort}
\usepackage{url}
\usepackage{deleq}
\usepackage{a4}
\usepackage{cite}
\usepackage{ntheorem}
\newcommand{\maclKzoz}{{\mathcal K}^{\bot_{g_0}}_0}

\newcommand{\calSJr}{{\cal S}_{(J_0,\rho_0)}}
\newcommand{\Ima}{\mbox{\rm Im}}
\newcommand{\Ker}{\mbox{\rm Ker}}

\newcommand{\riemgz}{g_0}

\renewcommand{\hbar}{{\overline \riemgz}}

\newcommand{\mcC}{{\mycal C}}

\newcommand{\nablash}{\nabla{\kern -.75 em
     \raise 1.5 true pt\hbox{{\bf/}}}\kern +.1 em}
\newcommand{\Deltash}{\Delta{\kern -.69 em
     \raise .2 true pt\hbox{{\bf/}}}\kern +.1 em}
\newcommand{\Rslash}{R{\kern -.60 em
     \raise 1.5 true pt\hbox{{\bf/}}}\kern +.1 em}

\newcommand{\Ric}{\operatorname{Ric}}

\newcommand{\hbound}{\mathring{\mathsf{H}}}
\newcommand{\cbord}{{\mathsf{C}}}

\newcommand{\chyp}{\mathcal C}
\newcommand{\hhyp}{\mathring{\mathcal H}}

\newcommand{\maclKzo}{{\mathcal K}^{\bot_g}_0}
\newcommand{\maclKz}{{\mathcal K}_0}
\newcommand{\hbord}{\hbound}%

\newcommand{\mcO}{{\mycal O}}

\newcommand{\mcU}{{\mycal U}}

\newcommand{\hyp}
{{\mycal S}}

\newcommand{\mcM}{{\mycal M}}
\newcommand{\mcK}{{\mycal K}}

\newcommand{\bea}{\begin{eqnarray}}
\newcommand{\beaa}{\begin{eqnarray*}}
\newcommand{\bean}{\begin{eqnarray}\nonumber}

\newcommand{\bel}[1]{\begin{equation}\label{#1}}
\newcommand{\beal}[1]{\begin{eqnarray}\label{#1}}
\newcommand{\beadl}[1]{\begin{deqarr}\label{#1}}
\newcommand{\eeadl}[1]{\arrlabel{#1}\end{deqarr}}
\newcommand{\eeal}[1]{\label{#1}\end{eqnarray}}
\newcommand{\eead}[1]{\end{deqarr}}
\newcommand{\eea}{\end{eqnarray}}
\newcommand{\eeaa}{\end{eqnarray*}}

\newcommand{\Ricc}{\mathrm{Ric}\,}

\newcommand{\Lpsi}{L^2_{\psi}}
\newcommand{\Lpsione}{\zH^1_{\phi,\psi}}
\newcommand{\Lpsitwo}{\zH^2_{\phi,\psi}}

\newcommand{\Lpsikg}[2]{\zH^{#1}_{\phi,\psi}(#2)}

\newcommand{\be}{\begin{equation}}
\newcommand{\ee}{\end{equation}}
\newcommand{\divr }{\mbox{\rm div}\,}
\newcommand{\tr}{\mbox{\rm tr}\,}

\newcommand{\eq}[1]{\eqref{#1}}
\newcommand{\Eq}[1]{Equation~(\ref{#1})}

\DeclareFontFamily{OT1}{rsfs}{}
\DeclareFontShape{OT1}{rsfs}{m}{n}{ <-7> rsfs5 <7-10> rsfs7 <10->
rsfs10}{} \DeclareMathAlphabet{\mycal}{OT1}{rsfs}{m}{n}

\newcommand{\zmcH }{\,\,\,\,\mathring{\!\!\!\!\mycal H}{}}

\usepackage{amsmath} 

\renewcommand{\theequation}{\thesection.\arabic{equation}}

\let\ssection=\section

\renewcommand{\section}{\setcounter{equation}{0}\ssection}

\newtheorem{defi}{\sc Definition\rm}[section]

\newtheorem{Theorem}[defi]{\sc Theorem\rm}

\newtheorem{prop}[defi]{\sc Proposition\!\rm}

\newtheorem{theor}[defi]{\sc Theorem\rm}

\newtheorem{Proposition}[defi]{\sc Proposition\rm}

\newtheorem{lem}[defi]{\sc Lemma\!\rm}
\newtheorem{Lemma}[defi]{\sc Lemma\!\rm}

\newtheorem{cor}[defi]{\sc Corollary\!\rm}
\newtheorem{Corollary}[defi]{\sc Corollary\!\rm}
\theorembodyfont{\upshape}
\newtheorem{Remark}[defi]{{\sc Remark}\rm}

\newtheorem{example}[defi]{{\sc Example}\rm}

\newtheorem{remark}[defi]{{\sc Remark}\rm}

\newcommand{\qed}{\hfill $\Box$\bigskip}

\newcommand{\proof}{\noindent {\sc Proof:\ }}

\def \Reel{\mathbb{R}}

\def \R {\Reel}

\def \Nat{\mathbb{N}}
\def \Z{\mathbb{Z}}
\def \N {\Nat}

\newcommand{\bM}{\,\overline{\!M}}

\newcommand{\zHkpp}{\zHk_{\phi,\psi}}
\newcommand{\Hkpp}{H^k_{\phi,\psi}}
\newcommand{\zHk}{\zH^k}
\newcommand{\zH}{\mathring{H}}

\newcounter{mnotecount}[section]

\newcommand{\rmnote}[1]{}

\newcommand{\loc}{{\textrm{loc}}}

\begin{document}
\title{Manifold structures for sets of solutions of the general relativistic constraint equations}
\author{
Piotr T. Chru\'{s}ciel\thanks{Visiting Scientist. Permanent
address: D\'epartement de Math\'ematiques, Facult\'e des Sciences,
Parc de Grandmont, F37200 Tours, France. Partially supported by a
Polish Research Committee grant 2 P03B 073 24, by the Erwin
Schr\"odinger Institute, and by a travel grant from the Vienna
City Council; email \protect\url{
piotr@gargan.math.univ-tours.fr}, URL \protect\url{
www.phys.univ-tours.fr/}$\sim$\protect\url{piotr}}\\
Erwann Delay\thanks{Visiting Scientist. Permanent address:
D\'epartement de Math\'ematiques, Facult\'e des Sciences, rue
Louis Pasteur, F84000 Avignon, France. Partially supported by the
ACI program of the French Ministry of Research and by the Erwin
Schr\"odinger Institute; email
  \protect\url{delay@gargan.math.univ-tours.fr}}
      \\
Erwin Schr\"odinger Institute\\ Vienna, Austria }
\date{}

\maketitle


\begin{abstract}
We construct  manifold structures on various sets of solutions of
the general relativistic initial data sets.
\\
\\
{AMS 83C05, 58J99; PACS 04.20.Cv}
\end{abstract}

\section{Introduction}

A natural question that arises in general relativity is whether
some sets of solutions of the, say vacuum, constraint equations
carry a manifold structure. For example, it is useful to have a
Banach manifold structure on the set of asymptotically flat
solutions of the constraint equations when trying to minimise the
ADM mass~\cite{Bartnik:phase,BrillDeserFadeev68,bartnik:qlm}.
Appropriate manifold structures allow one to use tools such as the
Smale-Sard theorem, or the Baire category theorem,  when
discussing genericity of some properties of solutions of the
Einstein equations. The existence of a \emph{Fr\'echet} manifold
structure\footnote{As pointed out to us by R.~Bartnik,  the
argument of~\cite{FischerMarsdenHI} leads to a loss of regularity
of the metric, forcing one to work with $C^\infty$ objects as far
as the manifold structure is concerned.} for (a subset of) the set
of solutions of the vacuum constraint equations on a compact
manifold is a consequence of linearisation stability studies of
Fischer, Marsden and Moncrief, (see~\cite{FischerMarsdenHI,key355}
and references therein, compare Theorem~\ref{TFrechet} below). The
results there have also been studied in the context of
asymptotically flat initial data sets on a manifold with compact
interior in~\cite{AnderssonJGP}.

While the above results might be satisfactory for several
purposes, they do \emph{not} lead to a \emph{Banach} {manifold} of
solutions. In finite dimension there is no need to introduce a
distinction between Banach, or Hilbert, or Fr\'echet manifold
structure; however, the differences are significant in infinite
dimension, because some facts which are true in Hilbert spaces are
not necessarily true in all Banach spaces. Similarly some
properties of Banach spaces do not carry over to Fr\'echet spaces.
(The reader is referred
to~\cite{Palais:analysis,Eells:analysis,KrieglMichor,Lang} for
analysis on infinite dimensional manifolds.)

In the asymptotically flat case an alternative method has been
developed by R.~Bartnik\footnote{R.~Bartnik, in preparation. Some
similar ideas have also been considered by L.~Andersson (private
communication).} for constructing a Hilbert manifold structure on
the space of solutions of the vacuum constraints, essentially
based on the conformal method. It uses  a (weighted) $H^{k}\times
H^{k+1}$ topology, $k\ge 1$, on the space of $(K,g)$'s, and it is
clear that the method generalises to certain other settings of
interest. Such spaces are well adapted to the evolution problem,
at least for $k$ large enough. However, in that method one does
not have the flexibility in controlling boundary or asymptotic
behavior which is provided by the
Corvino-Schoen~\cite{CorvinoSchoenprep,ChDelay} version of the
Fischer-Marsden-Moncrief approach.

 The purpose of this note is to show that a {\em Banach}
manifold structure can be obtained by a variation of the
Fischer-Marsden-Moncrief-Corvino-Schoen method. It turns out that
a direct application of the standard \emph{a priori}
estimates~\cite{Morrey} for the construction below does not seem
to lead to  a manifold structure based on Sobolev spaces, which
would have  been more natural for the evolution problem, and which
would have led to a \emph{Hilbert} manifold structure.  Instead,
\emph{e.g.}\/ on compact manifolds without boundary, we obtain a
manifold modeled on (a subset of) the space $C^{k,\alpha}\times
C^{k,\alpha}$, $k\ge 4$, $\alpha\in (0,1)$ of symmetric tensors.
 This appears
somewhat surprising at first sight, as a natural set-up for the
evolution problem (regardless of the Sobolev \emph{vs} H\"older
space issue) might seem to be one where the differentiability of
the extrinsic curvature tensor $K$ is one order less than that of
the metric $g$. On the other hand, since $K$'s can be thought of
as variations of $g$'s, from a manifold structure point of view it
seems natural that the $K$'s live in a space with the same
differentiability as $g$. Whatever the natural space is, the
$C^{k,\alpha}\times C^{k,\alpha}$ topology or weighted versions
thereof are the ones which are obtained by the method here; this
is a rather unexpected consequence of our analysis in this paper.
As already pointed out, and as made clear in the applications
below, the  manifolds of initial data obtained here exhibit more
structure than what is obtained by the conformal method and its
variations.

In our construction of the manifold structure  we use a smoothing
device to recover the loss of regularity inherent to the
Fischer-Marsden-Moncrief approach. This allows us to work
consistently in spaces with finite differentiability, leading to
the Banach manifold structure described above. We use a general
approach of weighted spaces as in~\cite{ChDelay}, which allows a
simultaneous treatment of the compact case with or without
boundary, and of the asymptotically flat case, and of the
conformally compactifiable case, with families of different
topologies.

All the results presented below remain valid in the
\emph{time-symmetric} setting, $K=0$. This implies that all the
manifold structures presented here have their obvious counterparts
for the set of Riemannian metrics with prescribed scalar
curvature.

 This paper is organised as follows: In
Section~\ref{SFrechet} we review the framework of~\cite{ChDelay},
and we show how the results there can be used to define a local
Hilbert manifold of solutions near a given solution; the resulting
manifolds can not be patched together in general because of
insufficient regularity of the operators involved. In
Section~\ref{Smrfd} we present our basic regularisation procedure,
which turns out to  still be insufficient to provide a (global)
Hilbert manifold structure. In Section~\ref{SAwHs} we therefore
pass to an analysis in weighted H\"older spaces. We prove there
that (the KID-free part of) the level sets of the constraints map
are, globally, embedded submanifolds in  a Banach space, see
Theorem~\ref{submanifold}, under very general conditions on the
weights; this is the main result of the paper. In fact, we prove
that the level sets of the constraint map foliate (in a sense made
precise in Remark~\ref{Rfol} below) the KID-free part of the space
of all $(K,g)$'s. In Section~\ref{Sappli} we show that the
hypotheses made in Theorem~\ref{submanifold} are fulfilled on
compact manifolds with or without boundary, or on asymptotically
compactifiable manifolds, or on asymptotically flat manifolds. In
appendix~\ref{A} we prove a  lemma which provides a submanifold
structure in Banach spaces under rather general conditions, as
well as a foliation result. In Appendix~\ref{Sconv} we present two
regularisation procedures in weighted spaces, as needed in
applications of the submanifold Theorem~\ref{submanifold}. For the
convenience of the reader those results from~\cite{ChDelay} which
play a key role in the current construction have been presented in
detail, including some introductory comments borrowed
from~\cite{ChDelay} whenever useful for the clarity of the
argument.

\section{
The construction}\label{SFrechet}

 Let $$ \mcC(K,g):=(J(K,g),\rho(K,g))$$ be the general relativistic {\em constraints map}:
  \be \label{1} \left(
\begin{array}{c}
J\\
  \\
\rho
\end{array}
\right) (K,g):= \left(
\begin{array}{l}
2(-\nabla^jK_{ij}+\nabla_i\;\tr  K)\\
  \\
R(g)-|K|^2 + (\tr  K)^2-2\Lambda
\end{array}
\right)\;, \ee  where $\Lambda$ is the cosmological constant. (The
function $c^4\rho/16\pi G $ is  the energy density of the matter
fields, while $c^4J/16\pi G $ is the energy-momentum flux vector.)
The general relativistic constraint equations are $\mcC(K,g)=0$,
whatever the space-dimension $n$. As those equations  are trivial
in space-dimension zero and one, in the remainder of this work we
shall assume that $n\ge 2$.

Let $h=\delta g$ and $Q=\delta K$, the linearisation $ P_{(K,g)}$
of the constraints map at $(K,g)$ reads \be \label{2}
P_{(K,g)}(Q,h)= \left(
\begin{array}{l}
-K^{pq}\nabla_i h_{pq}+K^q{}_i(2\nabla^j h_{qj}-\nabla_q h^l{}_l)\\
\;\;\;\;\;\;\;-2\nabla^jQ_{ij}+2\nabla_i\;\tr  Q
-2(\nabla_iK^{pq}-\nabla^qK^p{}_i)h_{pq}\\
  \\
-\Delta(\tr  h)+\divr    \divr    h-\langle h,\Ricc(g)\rangle +2K^{pl}K^q{}_l h_{pq}\\
\;\;\;\;\;\;\;-2\langle K,Q\rangle +2\tr  K(-\langle h,K\rangle
+\tr Q)
\end{array}
\right)\;. \ee

Recall that a KID is defined as a solution $(N,Y)$ of the set of
equations $P_{(K,g)}^*(Y,N)=0$, where $P_{(K,g)}^*$ is the formal
adjoint of $P_{(K,g)}$:\be \label{4} P_{(K,g)}^*(Y,N)= \left(
\begin{array}{l}
2(\nabla_{(i}Y_{j)}-\nabla^lY_l g_{ij}-K_{ij}N+\tr K\; N g_{ij})\\
 \\
\nabla^lY_l K_{ij}-2K^l{}_{(i}\nabla_{j)}Y_l+
K^q{}_l\nabla_qY^lg_{ij}-\Delta N g_{ij}+\nabla_i\nabla_j N\\
\; +(\nabla^{p}K_{lp}g_{ij}-\nabla_lK_{ij})Y^l-N \Ricc(g)_{ij}
+2NK^l{}_iK_{jl}-2N (\tr \;K) K_{ij}
\end{array}
\right) \;.\ee  We shall denote by $\mcK(\Omega)$ the set of KIDs
defined  on an open set $\Omega$.
In vacuum space-times $(\mcM,g)$ KIDs on a spacelike hypersurface
$\Omega$ are in one-to-one correspondence with Killing vectors of
$g$~\cite{MR50:15836} on the domain of dependence of $\Omega$. A
similar statement holds in electro-vacuum for appropriately
invariant initial data for the electromagnetic field, the reader
is referred to~\cite{ChBeigKIDs} for some comments about general
matter fields.

Following~\cite{ChDelay}, we will be using weighted function
spaces defined as follows. Let $\phi$ and $\psi$ be two smooth
strictly positive functions\footnote{\label{footmanif}We use the
analysts' convention that a manifold $M$ is always open; thus a
manifold $M$ with non-empty boundary $\partial M$ does not contain
its boundary; instead, $\bM:= M \cup
\partial M$ is a manifold with boundary in the differential
geometric sense. Unless explicitly specified otherwise \emph{no}
conditions on $M$ are made  --- \emph{e.g.}\/ that $\partial M$,
if non-empty, is compact
--- except that $M$ is a smooth manifold; similarly no conditions
\emph{e.g.}\/ on completeness of $(M,g)$, or on its radius of
injectivity, are made.} on $M$.  For $k\in \Nat$ let $\Hkpp(g) $
be the space of $H^k_\loc$ functions or tensor fields such that
the norm\footnote{The reader is referred to
\cite{Aubin,Aubin76,Hebey} for a discussion of Sobolev spaces on
Riemannian manifolds.}
 \be \label{defHn}
 \|u\|_{\Hkpp (g)}:=
(\int_M(\sum_{i=0}^k \phi^{2i}|\nabla^{(i)}
u|^2_g)\psi^2d\mu_g)^{\frac{1}{2}} \ee is finite, where
$\nabla^{(i)}$ stands for the tensor $\underbrace{\nabla ...\nabla
}_{i \mbox{ \scriptsize times}}u$, with $\nabla$ --- the
Levi-Civita covariant derivative of $g$; we assume throughout that
the metric is at least $W^{1,\infty}_\loc$; higher
differentiability will be usually indicated whenever needed. For
$k\in \Nat$ we denote by $\zHkpp $ the closure in $\Hkpp$ of the
space of $H^k$ functions or tensors which are compactly (up to a
negligible set) supported in $M$, with the norm induced from
$\Hkpp$.
The $\zHkpp $'s are Hilbert spaces with the obvious scalar product
associated to the norm \eq{defHn}. We will also use the following
notation
$$
\quad \zHk:=\zHk  _{1,1}\;,\quad
L^2_{\psi}:=\zH^0_{1,\psi}=H^0_{1,\psi}\;,
$$ so that $L^2\equiv \zH^0:=\zH^0_{1,1}$. We  set
$$
W^{k,\infty}_{\phi}:=\{u\in W^{k,\infty}_{\loc} \mbox{ such that }
\phi^i|\nabla^{(i)}u|_g\in L^{\infty}\}\;,
$$
with the obvious norm, and with $\nabla^{(i)}u$ --- the
distributional derivatives of $u$.

For $\phi$ and $\varphi$  --- smooth strictly positive functions
on M, and for $k\in\N$ and $\alpha\in [0,1]$, we define
$C^{k,\alpha}_{\phi,\varphi}$ the space of $C^{k,\alpha}$
functions or tensor fields  for which the norm
$$
\begin{array}{l}
\|u\|_{C^{k,\alpha}_{\phi,\varphi}(g)}=\sup_{x\in
M}\sum_{i=0}^k\Big(
\|\varphi \phi^i \nabla^{(i)}u(x)\|_g\\
 \hspace{3cm}+\sup_{0\ne d_g(x,y)\le \phi(x)/2}\varphi(x) \phi^{i+\alpha}(x)\frac{\|
\nabla^{(i)}u(x)-\nabla^{(i)}u(y)\|_g}{d^\alpha_g(x,y)}\Big)
\end{array}$$ is finite.

We will only consider weight functions with the property that
there exists $\ell \in \N\cup\{\infty\}$ such
that\footnote{Conditions~\eqref{lcond} will typically impose
$\ell$ restrictions on the behavior of the metric and its
derivatives in the asymptotic regions; it is therefore essential
to allow $\ell<\infty$ if one does not wish to impose an infinite
number of such conditions.} for $0\le i< \ell$ we have
\begin{equation}\label{lcond}
 |\phi^{i-1}\nabla^{(i)}\phi|_g\leq C_{i}\;,\;\;\;
|\phi^{i}\psi^{-1}\nabla^{(i)}\psi|_g\leq C_{i}\;,\;\;\;
|\phi^{i}\varphi^{-1}\nabla^{(i)}\varphi|_g\leq C_{i}\;,
\end{equation}
for some constants $C_i$. The following situations seem to be of
main interest:
\begin{itemize}

\item If $M$ is compact without boundary we will use
$\phi=\psi=\varphi=1$.

\item If $\partial M$ is compact, smooth, and non-empty,
 we will use for $\phi=x$ a function which is a
defining function for the boundary, at least in a neighborhood of
the boundary; that is, any smooth non-negative function on $\bM$
such that $\partial M$ is precisely the zero-level set of $x$,
with $dx$ without zeros on $\partial M$. Then $\psi$ and $\varphi$
will be a power of $x$ on a neighborhood of $\partial M$.
Condition \eq{lcond} will hold for metrics which are smooth
up-to-boundary near $\partial M$.

\item If $M$ contains an asymptotically flat region, $\phi$ will behave as
$r$, while $\varphi$ and $\psi$ will behave as a power of $r$ in
the asymptotically flat region; \eq{lcond} will hold for a large
class of asymptotically flat metrics.

\item If $M$ contains a conformally compactifiable region, then in a neighborhood of the conformal
boundary  $\phi$  will be taken to be $1$, while $\psi$ will be a
power of the defining function of the conformal boundary.

\item Exponentially weighted versions of the above will also be
considered.
\end{itemize}
In all those situations one can obtain elliptic estimates in
weighted spaces for the equations considered here by covering and
scaling arguments together with the standard interior elliptic
estimates on compact sets ({\em cf.,
e.g.}\/~\cite{choquet-bruhat:christodoulou:elliptic,Bartnik:mass,AndElli,AndChDiss,GL,Lee:fredholm}).
We will refer to this as \emph{the scaling property}. More
precisely, we shall say that \emph{the scaling property} holds
(with respect to some weighted Sobolev spaces with weight
functions $\psi$ and $\phi$, and/or weighted H\"older spaces with
weight functions $\varphi$ and $\phi$, whichever ones are being
used will always be obvious from the context) if there exists a
covering of $M$ by a family of sets $\Omega_\alpha$, for $\alpha$
in some index set $I$, together with scaling transformations
$\phi_\alpha:\Omega_\alpha\to \hat\Omega_\alpha$ on each of the
sets $\Omega_\alpha$, such that the transformed fields $(\hat
K_\alpha,\hat g_\alpha)$ on $\Omega_\alpha$ are in\footnote{It is
conceivable that in some situations less \emph{a priori}
regularity on the $(\hat K_\alpha,\hat g_\alpha)$'s can be
assumed, but this is the setup which seems to play the most
important role in our paper; the reader should be able to adapt
the differentiability conditions to his needs if required.} in
$W^{3,\infty}(\hat \Omega_\alpha)\times W^{4,\infty}(\hat
\Omega_\alpha)$, and such that the usual interior elliptic
estimates on the $\hat \Omega_\alpha$'s can be pieced together to
a weighted estimate, such as \eq{erc}, for the original fields.
Some sufficient conditions for the scaling property are discussed
in~\cite[Appendix~B]{ChDelay}.  We note that the scaling
transformation of the fields on $\hat \Omega_\alpha$,
$(K,g)\to(\hat K_\alpha, \hat g_\alpha)$,  will typically consist
of a pull-back of the fields, accompanied perhaps by a constant
conformal rescaling.
  The ``scaling property" is a
condition {both} on the {metric} $g$, the {extrinsic curvature
tensor} $K$, and on the {weight functions} involved: indeed, both
the metric coefficients, the connection coefficients, as well as
their derivatives, \emph{etc.}, \/which appear in our equations
must have appropriate behavior under the above transformations so
that the required estimates can be established.
%

Conditions~\eq{lcond} guarantee the following:
\begin{lem} \label{L4.3} Let $k\in\Z$, $k\geq -2$. Suppose that\footnote{The local
differentiability conditions follow from the requirement that the
$k+$ fourth covariant derivatives of $N$ and the $k+$ third ones
of $Y$ can be defined in a distributional sense; both of those
conditions are fulfilled by a metric $g\in W^{k+3,\infty}_\loc$
--- the reader should note that the first covariant derivatives of
$N$ do not involve the Christoffel symbols of $g$ since $N$ is a
function.}  $g\in W^{k+3,\infty}_\loc$ and that \be \label{Richyp}
\Ricc(g)\in\phi^{-2}W^{k+2,\infty}_{\phi}(g),\ee \be \label{Khyp}
K\in \phi^{-1}W^{k+3,\infty}_{\phi}(g)\;. \ee If \eq{lcond} holds
with $0\le i\le k+2$, then the linear operators \bel{P*map}
P_{(K,g)}^*:  \zH^{k+3}_{\phi,\psi}(g)\times
\zH^{k+4}_{\phi,\psi}(g)\longrightarrow \phi^{-1}
\zH^{k+2}_{\phi,\psi}(g)\times \phi^{-2}
\zH^{k+2}_{\phi,\psi}(g),\quad k\ge -2\;, \ee\bel{Pmap}
P_{(K,g)}:\psi^2(\phi\zH^{k+2}_{\phi,\psi}(g)\times
\phi^2\zH^{k+2}_{\phi,\psi}(g)) \longrightarrow
\psi^2(\zH^{k+1}_{\phi,\psi}(g)\times \zH^{k}_{\phi,\psi}(g)),
\quad k\ge 0\;, \ee are well defined, and bounded.
\end{lem}

 The
following operator  is of interest in our context, \bel{uprightL}
{L}_{\phi,\psi}:=  \psi^{-2} P_{(K,g)}\Phi\psi^2 \Phi P_{(K,g)}^*
\;,\ee where $\Phi$ is defined by \be\label{DefPhi}
\Phi(x,y):=(\phi x,\phi^2 y)\;.\ee A useful inequality to make
things work is the following: \be \label{it1kerbis} C\|\Phi
P_{(K_0,g_0)}^*(Y, N)\|_{\Lpsi(g_0)} \geq
\|Y\|_{\Lpsi(g_0)}+\|N\|_{\Lpsione(g_0)}\;. \ee Let ${\maclKz }$
be kernel of
$$\Phi P_{(K_0,g_0)}^*:
\Lpsikg{1}{g_0}\times \Lpsikg{2}{g_0}\longrightarrow
{\Lpsi}(g_0)\times {\Lpsi}(g_0),$$ and let ${\maclKz
}^{\bot_{g_0}}$ be its $\Lpsi(g_0)\oplus\Lpsi(g_0)$-orthogonal. We
denote by $\pi_{\maclKzo }$ the $L^2_\psi(g)$ projection onto
$\maclKzo $. The following result, proved\footnote{Actually
in~\cite{ChDelay} the hypothesis is made that the cosmological
constant $\Lambda$ vanishes, which is not assumed here (compare
\eq{1}). The inclusion of a cosmological constant does not require
any modifications of the proofs there, insofar as the results
discussed here are concerned.} in~\cite{ChDelay}, is a weighted
equivalent of those
in~\cite{Corvino,CorvinoSchoen,CorvinoOberwolfach}
(compare~\cite{CorvinoSchoenprep}):

\begin{theor}[\protect{\cite[Theorem~3.6]{ChDelay}}]\label{Til1projbis} Let $k\ge 0$, $g_0\in W^{k+4,\infty}_\loc$,
suppose that \eq{lcond} holds with $0\le i\le 4+k$, that
$$ \Ricc(g_0)\in\phi^{-2}W^{k+2,\infty}_{\phi}(g_0)\;,\quad
K_0\in \phi^{-1}W^{k+3,\infty}_{\phi}(g_0)\;,
$$
and that the weights $\phi$ and $\psi$ have the \emph{scaling
property}. If there exists a compact set $\mcK\subset M$ such that
for all $\Lpsione(g_0)$ vector fields $Y$ and $\Lpsitwo(g_0)$
functions $N$, both supported in $M\setminus \mcK$, the inequality
\eq{it1kerbis} holds, then for all $(K,g)$ close to $(K_0,g_0)$ in
$\phi^{-1}W^{k+3,\infty}_\phi(g_0)\times W^{k+4,\infty}_\phi(g_0)$
norm, the map$$ \pi_{\maclKzo }{L}_{\phi,\psi}:{{\maclKz
}^{\bot_g}}\cap (\Lpsikg{k+3}{g}\times \Lpsikg{k+4}{g})
\longrightarrow {\maclKzo }\cap (\Lpsikg{k+1}{g}\times
\Lpsikg{k}{g})
$$
is an  isomorphism such that the norm of its inverse is bounded
independently of $(K,g)$.

 \qed
\end{theor}

We will see how to use this result, and certain variations
thereof, to obtain a manifold structure on various sets of
solutions of the vacuum constraint equations. More generally, one
obtains a manifold structure on the set of initial data with
$(J,\rho)$--fixed. Recall that a \emph{Banach manifold} is a
Hausdorff topological space $M$ such that for every $p\in M$ there
exists a neighborhood $\mcU_p\subset M$ of $p$ equipped with a
homeomorphism $\phi_p$ from $\mcU_p$ to an open subset $\mcO_p$ of
some Banach space $B_p$. The couple $(\mcU_p,\phi_p)$, and
sometimes simply the set $\mcU_p$, will be called a
\emph{coordinate patch}. On overlapping coordinate patches
$\mcU_p$ and $\mcU_q$ the maps $\phi_p\circ \phi^{-1}_q$ are
supposed to be smooth diffeomorphisms from their domains
$\phi_q(\mcU_p\cap \mcU_q)\subset B_q$ to their images
$\phi_p(\mcU_p\cap \mcU_q)\subset B_p$. One can similarly define
the notion of a \emph{Hilbert} manifold, and of a \emph{Fr\'echet}
manifold.

A connected embedded submanifold of an open subset of a Banach
space is always a manifold modeled on any of its tangent spaces
(which are all necessarily diffeomorphic, compare
Corollary~\ref{Csubmanifold} below). We will actually prove that
the level sets of the constraint map form  embedded submanifolds
in such a space, which will provide the desired manifold
structure.

 We start with the following observation:

\begin{Proposition}[Local Hilbert manifold (and submanifold)]
\label{Plin} Under the hypotheses of Theorem~\ref{Til1projbis},
suppose that there are no KIDs:
\bel{nokids}\zH^{k+3}_{\phi,\psi}(g_0)\times
\zH^{k+4}_{\phi,\psi}(g_0)\supset{\maclKz   }=\{0\}\;.\ee
 Assume that the map
\be\label{et1projbisbis}
\begin{array}{c}
 \psi^2(\phi\zH^{k+2}_{\phi,\psi}(g_0)\times
\phi^2\zH^{k+2}_{\phi,\psi}(g_0)) \longrightarrow
\Lpsikg{k+1}{(g_0)}\times \Lpsikg{k}{(g_0)}\\
(\delta K,\delta g) \longmapsto \psi^{-2}\left\{\left(
\begin{array}{c}
J\\
\rho
\end{array}
\right)[(K_0,g_0)+(\delta K,\delta g)] -\left(
\begin{array}{c} J\\
\rho
\end{array}
\right)(K_0,g_0)\right\}
\end{array}
\ee is differentiable at zero. Then the set \bel{subhilbert}
\begin{array}{c}{\cal S}_0=\left\{(Q,h)\in
\psi^2(\phi \zH^{k+2}_{\phi,\psi}(g_0)\times \phi^2{\zH}
^{k+2}_{\phi,\psi}(g_0)),\right.\hspace{5cm} \\
\mbox{ }\hspace{2cm}
\left.(J,\rho)(K_0+Q,g_0+h)-(J,\rho)(K_0,g_0)=0
\right\}\end{array}\ee  is an embedded submanifold of $\psi^2(\phi
\zH^{k+2}_{\phi,\psi}(g_0)\times \phi^2{\zH}
^{k+2}_{\phi,\psi}(g_0))$ in a neighborhood of zero.
\end{Proposition}

\proof A rather general justification is provided by
Lemma~\ref{LA1}, Appendix~\ref{A}, by setting
$u=Du(0)=\psi^2\Phi^2P_{(K_0,g_0)}^*$,
$v=\psi^{-2}[(J,\rho)(K_0+.,g_0+.)-(J,\rho)(K_0,g_0)]$,
$Dv(0)=\psi^{-2}P_{(K_0,g_0)}$,
$E=\zH^{k+3}_{\phi,\psi}(g_0)\times\zH^{k+4}_{\phi,\psi}(g_0)$,
$F=\psi^2\Phi
(\zH^{k+2}_{\phi,\psi}(g_0)\times\zH^{k+2}_{\phi,\psi}(g_0))$, and
$G=\zH^{k+1}_{\phi,\psi}(g_0)\times\zH^{k}_{\phi,\psi}(g_0)$. The
linear map $u$ is continuous by Lemma~\eq{L4.3}, hence
differentiable, while $v$ is differentiable by hypothesis. The
required isomorphism property for $Dv(0)\circ Du(0)$ follows from
Theorem~\ref{Til1projbis}. \qed

\begin{example}
\label{Ex1}To motivate and illustrate the results so far, consider
the case of a compact manifold $M$ without boundary; other
examples of interest will be treated in detail in
Section~\ref{Sappli}. We choose $\phi=\psi\equiv 1$, so the spaces
involved are standard Sobolev spaces. As $M$ is compact we can
take $\mcK=M$ so that condition \eq{it1kerbis} is trivially
satisfied. The smoothness hypotheses on the map \eq{et1projbis}
are satisfied by standard calculus in Sobolev spaces. If
$(K_0,g_0)$ is a $C^{k+4}\times  C^{k+3}$ initial data set without
KIDs, Proposition~\ref{Plin} provides a manifold of $H^{k+2}\times
H^{k+2}$ solutions of the constraint equations passing through
$(K_0,g_0)$.
\end{example}

Example~\ref{Ex1} clearly exhibits an unfortunate
differentiability mismatch here which leads to an essential
obstruction when trying to glue together the coordinate patches
obtained so far, and which therefore prevents one from obtaining a
Hilbert manifold structure on the set of all solutions of the
constraints equations using the method above. In
Section~\ref{Sbms} we will show that a (different) manifold
structure can nevertheless be obtained using  the following
approach: consider a data set $(K_1,g_1)$ without KIDs in an
appropriate H\"older differentiability class, let $(\delta K_1,
\delta g_1)$ be a solution of the constraints with the same
differentiability class, and suppose that you can smooth out
$(K_1+\delta K_1,g_1+ \delta g_1)$ to a \emph{smooth} set $(K,g)$,
in a way consistent with the set-up of
Theorem~\ref{theor:projbis}. If one can solve the equation
\bel{et1proj2bnew} \left(
\begin{array}{c}
J\\
\rho
\end{array}
\right) ((K_0+\delta K,g_0+\delta g)+\psi^2\Phi^2
P^*_{(K,g)}(Y,N)) - \left(
\begin{array}{c}
J\\
\rho
\end{array}
\right) (K_0+\delta K,g_0+\delta g)=\left(
\begin{array}{c}
\delta J\\
\delta \rho
\end{array}
 \right)\;,\ee
 then one has a better chance of ending in a space with the original
 differentiability.

\section{A regularised problem}
\label{Smrfd}

The aim of this section is to implement the above in weighted
Sobolev spaces. Consider again a compact manifold without
boundary, let $(K,g)$ be of $C^{k+2,\alpha}\times C^{k+2,\alpha}$
differentiability class, and first regularise $(K,g)$ by the usual
convolution method to obtain a new smooth couple
$(K_\varepsilon,g_\varepsilon)$, then define
$$
P^*_\varepsilon:=P^*_{(K_\varepsilon,g_\varepsilon)}\;.
$$
 Roughly speaking, the  equation we will attempt to solve will be
\eq{et1proj2bnew} with $P^*_{(K,g)}$ there replaced by
$P^*_\varepsilon$. The idea is to solve that equation for
$(Y,N)\in C^{3,\alpha}\times C^{4,\alpha}$, and then use elliptic
regularity to get to $(Y,N)\in C^{k+3,\alpha}\times
C^{k+4,\alpha}$, obtaining the desired  extrinsic curvature and
metric in $C^{k+2,\alpha}\times C^{k+2,\alpha}$; we emphasise that
one would \emph{not} get that last differentiability without the
regularisation. Somewhat more generally, let $0\le k_0\le k-2$,
still in the compact boundaryless case one has the following
convergence property:
$$P^*_\varepsilon \to P^*\ \mbox{ in } \  L(\zH^{k_0+3}_{\phi,\psi}(g_0)\times
\zH^{k_0+4}_{\phi,\psi}(g_0),\phi^{-1}
\zH^{k_0+2}_{\phi,\psi}(g_0)\times \phi^{-2}
\zH^{k_0+2}_{\phi,\psi}(g_0))\;,
$$
which is precisely what is needed to make the arguments work. In
general we will therefore assume that we have a smoothing
operation $(K,g)\to (K_\varepsilon,g_\varepsilon)$ such that
\bel{Psmooth}(K_\varepsilon,g_\varepsilon) \to _{\varepsilon\to
0}(K,g) \ \mbox{in}\  \phi^{-1}W^{k_0+3,\infty}_\phi(g_0)\times
W^{k_0+4,\infty}_\phi(g_0)\;. \ee (In Appendix~\ref{Sconv} we give
conditions on the weight functions which guarantee that the
smoothing operation with the above properties exists.) This leads
to the following variation of Theorem~3.9 of~\cite{ChDelay}:

\begin{theor}\label{theor:projbis}Let $k\ge 0$, $g_0\in W^{k+4,\infty}_\loc$,
assume that \eq{Psmooth} holds with $k_0=k$ and suppose that
\eq{lcond} holds with $0\le i\le 4+k$. Assume moreover that
$$ \Ricc(g_0)\in\phi^{-2}W^{k+2,\infty}_{\phi}(g_0)\;,\quad
K_0\in \phi^{-1}W^{k+3,\infty}_{\phi}(g_0)\;,
$$
and that the weights $\phi$ and $\psi$ have the \emph{scaling
property}. Suppose further that there exists a compact set
$\mcK\subset M$ such that for all $\Lpsione(g_0)$ vector fields
$Y$ and $\Lpsitwo(g_0)$ functions $N$, both supported in
$M\setminus \mcK$, the inequality \eq{it1kerbis} holds.
 If the weights are such that the map
\be\label{et1projbis}
\begin{array}{c}
 {\maclKzo }\cap (\Lpsikg{k+3}{g}\times  \Lpsikg{k+4}{g})
\longrightarrow
{\maclKzo }\cap (\Lpsikg{k+1}{g}\times \Lpsikg{k}{g})\\
(Y,N) \longmapsto \pi_{\maclKzo }\psi^{-2}\left\{\left(
\begin{array}{c}
J\\
\rho
\end{array}
\right)[(K,g)+\psi^2\Phi^2 P_\varepsilon^*(Y,N)] -\left(
\begin{array}{c} J\\
\rho
\end{array}
\right)(K,g)\right\}
\end{array}
\ee is differentiable in a neighborhood of zero, then it is
bijective in a (perhaps smaller) neighborhood of zero, for all
$\varepsilon$ small enough. In other words, for sufficiently small
$\varepsilon$, there exists $\delta>0$ such that for all $(K,g)$
close to $(K_0,g_0)$ in $\phi^{-1}W^{k+3,\infty}_\phi(g_0)\times
W^{k+4,\infty}_\phi(g_0)$, and for all pairs $(\delta J,
\delta\rho)\in  \psi^2\Big(\Lpsikg{k+1}{g}\times
\Lpsikg{k}{g}\Big)$ with norm less than $\delta$,
 there exists a solution
 \bel{solform}(\delta
K,\delta g)=\Phi\psi^2\Phi P_\varepsilon^*(Y,N) \in\psi^2(\phi
H^{k+2}_{\phi,\psi}(g)\times \phi^2 H^{k+2}_{\phi,\psi}(g))\;,\ee
close to zero, of the equation \bel{et1proj2} \pi_{\maclKzo
}\psi^{-2} \left\{\left(
\begin{array}{c}
J\\
\rho
\end{array}
\right) (K+\delta K,g+\delta g) - \left(
\begin{array}{c}
J\\
\rho
\end{array}
\right) (K,g) \right\}= \pi_{\maclKzo }\psi^{-2}\left(
\begin{array}{c}
\delta J\\
\delta \rho
\end{array}
\right) \;.\ee The solutions of the form \eq{solform} with
sufficiently small norm are unique.
\end{theor}

\proof
 Instead of \eq{et1proj2} we consider the projection of the equation
\bel{et1proj2bnewb} \left(
\begin{array}{c}
J\\
\rho
\end{array}
\right) ((K,g)+\psi^2\Phi^2 P^*_\varepsilon(Y,N)) - \left(
\begin{array}{c}
J\\
\rho
\end{array}
\right) (K,g)=\left(
\begin{array}{c}
\delta J\\
\delta \rho
\end{array}
 \right)\;,\ee
 If $\varepsilon$ is small enough then $(K_\varepsilon,g_\varepsilon)$ is close to
$(K,g)$ in $\phi^{-1}W^{k_0+3,\infty}_\phi(g_0)\times
W^{k_0+4,\infty}_\phi(g_0)$ thus close to $(K_0,g_0)$ in the same
space. Because of the high differentiability threshold assumed all
the coefficients in the equations are in $L^\infty_\loc$, and it
is easy to check that the operator $P^*_\varepsilon$ converges to
$P^*$ when $\varepsilon$ goes to zero  in such a way that the
estimates in Appendix~G of~\cite{ChDelay} remain uniform. It
follows that for $\varepsilon$ small enough $P^*_\varepsilon$ can
be used in place of $P^*$ to define the right inverse needed in
the arguments of Appendix~G of~\cite{ChDelay}. \qed

One would like to use elliptic estimate arguments to show that if
$(K,g)$ is smooth, and if $(\delta J, \delta \rho)$ is smooth,
then the solution is smooth. We have not been able to implement
such an argument in the spaces used above because of poor
differentiability of the coefficients of the equations. This has
the effect that the size of the neighborhood for which the theorem
applies might depend upon $k$. This problem will be sidetracked by
working in weighted H\"older spaces.

\section{Analysis in weighted H\"older spaces}\label{SAwHs}

Before passing to an analysis of the regularised equation
\eq{et1proj2bnew},  let us show that the results established
in~\cite{ChDelay} can be used to obtain a Fr\'echet manifold of
smooth solutions of constraint equations without KIDs.

In order to obtain a coherent set-up in weighted H\"older spaces
we will need to impose some more conditions on the weight
functions $\phi$, $\varphi$, and $\psi$:
\begin{enumerate} \item First,  note that \eq{lcond} can be
rewritten as $\phi\in C^{\ell-1}_{\phi,\phi^{-1}}$, $\psi\in
C^{\ell-1}_{\phi,\psi^{-1}}$, $\varphi\in
C^{\ell-1}_{\phi,\varphi^{-1}}$. When dealing with H\"older spaces
one also needs to assume H\"older continuity of the derivative
weights, so (renaming $\ell-1$ to $\ell$) we will assume:
\bel{threecB}\phi\in C^{\ell,\alpha}_{\phi,\phi^{-1}}\;, \quad
\psi\in C^{\ell,\alpha}_{\phi,\psi^{-1}}\;, \quad \varphi\in
C^{\ell,\alpha}_{\phi,\varphi^{-1}}\;.\ee
\item As discussed in \cite[Appendix~B]{ChDelay},  the following conditions
are useful for deriving the scaling property: Let us denote by
$B_p $ the open ball of center $p$ with radius $\phi(p) /2$. We
assume that there exist  constants $C_1,C_2,C_3>0$ such that for
all $p\in M $ and all $y\in B_p $, we have \beal{scalprop1}
C_1^{-1}\phi(p) \leq \phi (y)\leq C_1\phi(p) \;,\\
\label{scalprop2}
C_2^{-1}\varphi(p)\leq \varphi (y)\leq C_2\varphi(p)\;,\\
\label{scalprop3} C_3^{-1}\psi(p)\leq \psi (y)\leq C_3\psi(p)\;.
\eea
\item Since the tool to handle non-linearities in this paper is the inverse function  theorem,
we need to make sure that the changes in the initial data are
small as compared to the data themselves. A necessary condition
for that is that the new metric be uniformly equivalent to the
original one. One way of ensuring this is \bel{threedB}
\psi^2\phi^2C^{k,\alpha}_{\phi,\varphi}(g_0)\subset
C^{k,\alpha}_{\phi,1}(g_0) \;.\ee This will hold under the
following condition:
\begin{Proposition}
\label{Linclus} The
inequality\bel{threeaB}\psi^2\phi^2\varphi^{-1}\leq C\;.\ee
 implies \eq{threedB}.
\end{Proposition} In order to
check this the reader might wish to prove first that the
conditions imposed so far imply that
\begin{Lemma}\label{Lproduit}
If $u\in C^{k,\alpha}_{\phi,\varphi_1}(g)$ and $v\in
C^{k,\alpha}_{\phi,\varphi_2}(g)$, with one of the $\varphi_a$'s
satisfying
 \eq{scalprop2} and $\phi$ satisfying \eq{threecB} with $\ell\ge k$, then $uv\in
C^{k,\alpha}_{\phi,\varphi_1\varphi_2}(g)$.
\end{Lemma}
Lemma~\ref{Lproduit} can be used to show an equivalent of
Lemma~\ref{L4.3} in weighted H\"older spaces.
\end{enumerate}

Clearly all those conditions are fulfilled when
$\phi=\varphi=\psi=1$; they will also be fulfilled in the other
examples of interest discussed in Section~\ref{Sappli}.

To proceed further some terminology will be needed:

\begin{defi}\label{def:regul}
We will say that an operator $L$ from $\zH^{3}_{\phi,\psi}\times
\zH^{4}_{\phi,\psi}$ to $\zH^{1}_{\phi,\psi}\times
\zH^{0}_{\phi,\psi}$ satisfies the {\em weighted elliptic
regularity condition} if there exists a constant $C$ such that for
all $(Y,N)$ in $\zH^{3}_{\phi,\psi}\times \zH^{4}_{\phi,\psi}$
satisfying $L(Y,N)\in C^{k+1,\alpha}_{\phi,\varphi}\times
C^{k,\alpha}_{\phi,\varphi}$ we have $(Y,N)\in
C^{k+3,\alpha}_{\phi,\varphi}\times C^{k+4,\alpha}_{\phi,\varphi}$
with \bel{erc} \|(Y,N)\|_{C^{k+3,\alpha}_{\phi,\varphi}\times
C^{k+4,\alpha}_{\phi,\varphi}}\leq C \left(\|L(Y,N)\|_{
C^{k+1,\alpha}_{\phi,\varphi}\times C^{k,\alpha}_{\phi,\varphi}}
+\|(Y,N)\|_{H^{3}_{\phi,\psi}\times H^{4}_{\phi,\psi}}\right). \ee
\end{defi}

When ${L}_{\phi,\psi}$ defined in \eq{uprightL} satisfies the
weighted elliptic regularity condition one has the following:

\begin{prop}[Proposition~3.16 of~\cite{ChDelay}]\label{regularitebis}
Let $k\in \N$, $0<\alpha<1$, assume that \eq{threecB} with $\ell
\ge k+4$ holds, and that \eq{scalprop1}-\eq{scalprop3} and
\eq{threeaB} hold. In addition to the hypotheses of
Theorem~\ref{theor:projbis} with $\varepsilon=0$, suppose that
$g_0\in C^{k+4,\alpha}$, and that
$$ \Ricc(g_0)\in\phi^{-2}C^{k+2,\alpha}_{\phi,1}(g_0)\;,\quad
K_0\in \phi^{-1}C^{k+3,\alpha}_{\phi,1}(g_0)\;.
$$
We further assume that the weights $\phi$, $\varphi$ and $\psi$
have the \emph{scaling property}.  Suppose, next, that we have the
continuous inclusions \bel{poidscond2}\psi^2 \phi^2
C^{i,\alpha}_{\phi,\varphi^2}(g)\subset
\Lpsikg{i}{g} \ee for $i=k,k+1$, with the inclusion norms
uniformly bounded for $g$ close to $g_0$ in
$C^{k+4,\alpha}_{\phi,1}(g_0)$. Assume finally that ${
L}_{\phi,\psi}(K,g)$ satisfies the weighted elliptic regularity
condition, with a uniform constant $C$ in \eq{erc} for $(K,g)$
close to $(K_0,g_0)$ in $
\phi^{-1}C^{k+3,\alpha}_{\phi,1}(g_0)\times
C^{k+4,\alpha}_{\phi,1}(g_0)
$. If the source $(\delta J, \delta\rho)$ is in
$\psi^2(\zH^{1}_{\phi,\psi}(g)\times \zH^{0}_{\phi,\psi}(g))\cap
\psi^2(C^{k+1,\alpha}_{\phi,\varphi}(g)\times
C^{k,\alpha}_{\phi,\varphi}(g))$, with sufficiently small norm,
then the solution obtained  in  Theorem \ref{theor:projbis} with
$\varepsilon=0$ is in
$$\psi^2(\phi\zH^{2}_{\phi,\psi}(g)\times
\phi^2\zH^{2}_{\phi,\psi}(g))\cap\psi^2(\phi
C^{k+2,\alpha}_{\phi,\varphi}(g)\times
\phi^2C^{k+2,\alpha}_{\phi,\varphi}(g))\;.$$

\qed
\end{prop}

Proposition~\ref{regularitebis} gives existence of H\"older
continuous solutions. We can apply the usual bootstrap arguments
to those solutions to obtain smoothness, when all the objects at
hand are smooth (compare the proof of Theorem~\ref{Tbetter3}
below):

\begin{prop}[Proposition~3.17
of~\cite{ChDelay}]\label{regularite3} Let $k\in \N$, $\alpha\in
(0,1)$, assume that \eq{threecB} with $\ell \ge k+4$ holds, and
that \eq{scalprop1}-\eq{scalprop3} and \eq{threeaB} hold. Suppose
moreover  that the scaling property holds. Assume that $(K,g)\in
C^{k+3,\alpha}\times C^{k+4,\alpha} $ and $(Y,N)\in
C^{3,\alpha}_{\phi,\varphi}(g)\times
C^{4,\alpha}_{\phi,\varphi}(g) $. If
\be\label{et1projregbis4}\hspace{-0.3cm} \left(
\begin{array}{c}
J\\
\rho
\end{array}
\right)[(K,g)+\psi^2 \Phi^2P^*_{(K,g)}(Y,N)]-\left(
\begin{array}{c}
J\\
\rho
\end{array}
\right)[(K,g)] \in \psi^2(C^{k+1,\alpha}_{\phi,\varphi}(g)\times
C^{k,\alpha}_{\phi,\varphi}(g))\;, \ee
 then
$(Y,N)\in C^{k+3,\alpha}_{\phi,\varphi}(g)\times
C^{k+4,\alpha}_{\phi,\varphi}(g))$, thus \bel{finreg}(\delta
K,\delta g)\in \psi^2(\phi C^{k+2,\alpha}_{\phi,\varphi}(g)\times
\phi^2C^{k+2,\alpha}_{\phi,\varphi}(g))\;.\ee
\end{prop}

\noindent{\sc Example~\ref{Ex1} (continued):} Applying the last
two propositions to the setup of Example~\ref{Ex1} one finds that
smooth solutions of the linearised constraint equations correspond
to smooth solutions of the full non-linear constraint equations.
This leads then to a Fr\'echet manifold of smooth solutions near
every smooth solution. The argument at the end of proof of
Theorem~\ref{submanifold} justifies the isomorphism property on
the overlaps of the coordinate charts, and we have therefore
obtained the Fischer-Marsden-Moncrief result:

\begin{Theorem}[Fischer, Marsden, Moncrief~\cite{key355}]
\label{TFrechet} Let $M$ be a compact manifold with boundary. Then
the level sets $\cal S$ of the constraints map form a submanifold
of the set of smooth $(K,g)$'s at all $(K,g)$ which have no KIDs.
Each connected component ${\cal S}_0$ thereof is a Fr\'echet
manifold modeled on $\Ker P_{(K,g)} \subset C^\infty\times
C^\infty$, where $(K,g)$ is an arbitrary element of ${\cal S}_0$.
\end{Theorem}

Specialising all the considerations so far to the case $K\equiv
Y\equiv 0$ one recovers a theorem essentially due to Fischer and
Marsden:

\begin{Theorem}[Fischer, Marsden~\cite{FischerMarsdenScCurv}]
\label{TFrechet2} Let $M$ be a compact manifold with boundary.
Then the level sets of the scalar curvature functional on the
space of smooth metrics form a submanifold at all $g$ which do not
correspond to the space-part of some static solution of the vacuum
Einstein equations with a cosmological constant. Each connected
component ${\cal S}_0$ thereof is a Fr\'echet manifold modeled on
the kernel of $DR$, as calculated at some arbitrarily chosen
metric $g\in {\cal S}_0$.
\end{Theorem}

 The argument of the proof of Theorem~\ref{theor:projbis}
also establishes:

\begin{prop}\label{better2} Under the conditions of
Proposition~\ref{regularitebis},
 assume that  \eq{Psmooth} holds,  and
suppose that $\psi^{-2}P\psi^{2}\Phi^2P^*_\varepsilon$ satisfies
the weighted elliptic regularity condition, with the constant $C$
in \eq{erc} being uniform for $(K,g)$ close to $(K_0,g_0)$ in
$\phi^{-1}C^{k_0+3,\alpha}_{\phi,1}(g_0)\times
C^{k_0+4,\alpha}_{\phi,1}(g_0)$ and $\varepsilon$ small enough.
Then Proposition~\ref{regularitebis} remains valid with $k$
replaced by $k_0$ and  $P^*$ replaced by $P^*_\varepsilon$ when
$\varepsilon$ is small enough.
\end{prop}

Our first main result is the equivalent of Proposition
\ref{regularite3} with less regularity conditions on $(K,g)$:

\begin{theor}\label{Tbetter3}
Let $k\in \N$, $k\geq2$, $\alpha\in (0,1)$, assume that
\eq{threecB} with $\ell \ge k+4$ holds, and that
\eq{scalprop1}-\eq{scalprop3} and \eq{threeaB} hold. Under
\eq{Psmooth}, suppose that  the scaling condition holds for
$\varepsilon>0$.
Assume that $(K,g)\in C^{k+2,\alpha}
\times C^{k+2,\alpha} $ 
 and $(Y,N)\in C^{3,\alpha}_{\phi,\varphi}(g)\times
C^{4,\alpha}_{\phi,\varphi}(g) $. For $\varepsilon$ small, if
\be\label{betterequa}\hspace{-0.3cm} \left(
\begin{array}{c}
J\\
\rho
\end{array}
\right)[(K,g)+\psi^2 \Phi^2P_\varepsilon^*(Y,N)]-\left(
\begin{array}{c}
J\\
\rho
\end{array}
\right)[(K,g)] \in \psi^2(C^{k+1,\alpha}_{\phi,\varphi}(g)\times
C^{k,\alpha}_{\phi,\varphi}(g))\;, \ee
 then
$(Y,N)\in C^{k+3,\alpha}_{\phi,\varphi}(g_\varepsilon)\times
C^{k+4,\alpha}_{\phi,\varphi}(g_\varepsilon)$, thus $$(\delta
K,\delta g)\in \psi^2\left(\phi
C^{k+2,\alpha}_{\phi,\varphi}(g)\times
\phi^2C^{k+2,\alpha}_{\phi,\varphi}(g)\right)\;.$$
\end{theor}

\begin{Remark} The appearance of $\varepsilon$  in the claim that
$(Y,N)\in C^{k+3,\alpha}_{\phi,\varphi}(g_\varepsilon)\times
C^{k+4,\alpha}_{\phi,\varphi}(g_\varepsilon) $ is due to the fact
that $g$ is \emph{a priori} not sufficiently differentiable to be
able to define  spaces such as
$C^{k+3,\alpha}_{\phi,\varphi}(g_\varepsilon)$. Any fixed metric
uniformly equivalent to $g$, with appropriate weighted
differentiability properties, could be used instead of
$g_\epsilon$.
\end{Remark}

\proof It suffices to rewrite the rescaled non-linear elliptic
equation \eq{et1projregbis4} for $(Y,N)$ as a linear elliptic
equation for $(Y,N)$ and freeze coefficients (depending on
$(K+\delta K, g+\delta g)$ hence on $(Y,N)$). The  interior
H\"older estimates~\cite[Theorem~6.2.5, p.~223]{Morrey} on the
sets $\hat \Omega_\alpha$ appearing in the definition of scaling
property give the local regularity, and the scaling property gives
the global weighted regularity. \qed

\section{Banach manifold structure}\label{Sbms}
Throughout this  section the symbol $g_0$ denotes a fixed metric
with (local) regularity $C^{m+4,\alpha}$ on $M$.

For $k,l\in \{0,...,m+4\}$, $\alpha\in(0,1)$ and $g$ a metric in
$C^{\max(l,k),\alpha}$, we define the Banach space
$$
 \Lambda^{l,k,\alpha}_{\phi,\psi,\varphi}(g)= \zH^{l}_{\phi,\psi}(g)\cap C^{k,\alpha}_{\phi,\varphi}(g)\;,$$
 equipped with a norm being the sum of the two norms.
 (It should be clear from \eq{finreg} that this is the topology
which one needs to use on the space of the metrics when using the
methods described above.) When $k\in\{0,...,m+2\}$ and when
\eq{threedB} hold, we define the following {\it open} subset of
$\psi^2\phi^2\Lambda^{2,k+2,\alpha}_{\phi,\psi,\varphi}(g_0)$ of
symmetric two-covariant tensor fields on $M$:
$$
{\cal A}^{k+2,\alpha}_{\phi,\psi,\varphi}(g_0):= \left \{h \in
{\psi^2\phi^2\Lambda^{2,k+2,\alpha}_{\phi,\psi,\varphi}(g_0)}\;,\
g_0+h \mbox{ is a metric uniformly equivalent to }g_0\right\}\;.$$
We note the continuous inclusions: \bel{threebB}{\cal
A}^{k+2,\alpha}_{\phi,\psi,\varphi}(g_0)\subset
\psi^2\phi^2\Lambda^{2,k+2,\alpha}_{\phi,\psi,\varphi}(g_0)\subset
C^{k+2,\alpha}_{\phi,1}(g_0)\;\Big(\subset
W^{k+2,\infty}_\phi(g_0)\Big)\;.\ee
 We have the
\begin{lem}\label{samespace}
Let $k\in\{0,...,m+2\}$ and    $l\in\{0,...,k+2\}$. Assume
\eq{threecB}-\eq{threedB}.  Then for all $h\in{\cal
A}^{k+2,\alpha}_{\phi,\psi,\varphi}(g_0)$ we have
$$C^{l,\alpha}_{\phi,\varphi}(g_0+h)=C^{l,\alpha}_{\phi,\varphi}(g_0)\;,
\quad \zH^{l}_{\phi,\psi}(g_0+h)=\zH^{l}_{\phi,\psi}(g_0)\;,$$
with equivalent norms. In particular,
$$
{\cal A}^{k+2,\alpha}_{\phi,\psi,\varphi}(g_0+h)={\cal
A}^{k+2,\alpha}_{\phi,\psi,\varphi}(g_0).$$
\end{lem}

\proof Let $g=g_0+h$, we define $T=\Gamma-\Gamma_0$, since $g$ is
uniformly equivalent to $g_0$ the usual formula for $T$ allows one
to estimate this by $C |\nabla_0 h|_{g_0}$.   By the middle
inclusion in \eq{threebB} we then have $T\in
\phi^{-1}C^{k+1,\alpha}_{\phi,1}(g_0)$. For a tensor $u$, we have
$\|u\|_{g_0}$ uniformly equivalent to $\|u\|_g$. For the
derivatives we write
$$\nabla
u=\nabla_0u+(\nabla-\nabla_0)u=\nabla_0u-Tu\;.$$ If $u\in
C^{l,\alpha}_{\phi,\varphi}(g_0)$ by the product
Lemma~\ref{Lproduit} we obtain $\phi \nabla u \in
C^{0,\alpha}_{\phi,\varphi}(g)$.  The higher derivatives follow by
induction. This shows that
$C^{l,\alpha}_{\phi,\varphi}(g_0)\subset
C^{l,\alpha}_{\phi,\varphi}(g)$.

We note that the  above implies that $-h\in
{\psi^2\phi^2C^{k+2,\alpha}_{\phi,\varphi}(g)}$, and the reverse
inclusion follows by symmetry.

The proof for the Sobolev spaces is identical.\qed

\begin{theor}\label{submanifold} Let $k\in \{2,...,m\}$ and
$\alpha\in(0,1)$, and \beal{RicK}
&\Ricc(g_0)\in\phi^{-2}C^{k,\alpha}_{\phi,1}(g_0)\;,\quad K_0\in
\phi^{-1}C^{k+2,\alpha}_{\phi,1}(g_0)\;.&\eea
Suppose that the scaling property  and the weighted regularity
condition hold, and that \eq{threecB}-\eq{scalprop3} together with
\eq{threeaB} are satisfied. Assume also that for
all\footnote{Actually it suffices to assume that this hypothesis
holds on $\calSJr$.} \bel{Qhc} (Q,h)\in \psi^2\phi
\Lambda^{2,k+2,\alpha}_{\phi,\psi,\varphi}(g_0)\times {\cal A}
^{k+2,\alpha}_{\phi,\psi,\varphi}(g_0)\ee there exists a compact
set $\mcK\subset M$ such that for all $\Lpsione(g_0)$ vector
fields $Y$ and $\Lpsitwo(g_0)$ functions $N$, both supported in
$M\setminus \mcK$, the inequality \eq{it1kerbis} holds with
$(K_0,g_0)$ there replaced by $(K_0+Q,g_0+h)$ . Suppose, further,
that for all $(Q,h)$ as in \eq{Qhc} the map \bel{cmap} (\delta K,
\delta g) \to (J,\rho)(K_0+Q+\delta K,g_0+h+\delta
g)-(J,\rho)(K_0+Q,g_0+h)\ee is differentiable from a neighborhood
of zero in $\psi^2\phi \zH^{2}_{\phi,\psi}(g_0)\times\psi^2\phi^2
\zH^{2}_{\phi,\psi}(g_0) $ to  $\psi^2
\zH^{1}_{\phi,\psi}(g_0)\times \psi^2 \zH^{0}_{\phi,\psi}(g_0)$.
 Consider any non-empty connected
component of the set of KID-free level-sets:\bel{sub}
\begin{array}{c}\calSJr=\left\{(Q,h)\in
\psi^2\phi \Lambda^{2,k+2,\alpha}_{\phi,\psi,\varphi}(g_0)\times
{\cal A}
^{k+2,\alpha}_{\phi,\psi,\varphi}(g_0),\right.\hspace{5cm} \\
\mbox{ }\hspace{2cm}\left.(J,\rho)(K_0+Q,g_0+h)=(J_0,\rho_0), \;{
\Ker }\;P^*_{K_0+Q,g_0+h}=\{0\}\right\}\end{array}\ee If there
exists a smoothing operation \eq{Psmooth} with $k_0=0$,
then  $\calSJr$ is an embedded submanifold of $\psi^2\phi
\Lambda^{2,k+2,\alpha}_{\phi,\psi,\varphi}(g_0)\times {\cal A}
^{k+2,\alpha}_{\phi,\psi,\varphi}(g_0)$.
\end{theor}
\begin{remark}
Differentiability of the map \eq{cmap} in weighted Sobolev spaces
typically requires $k>n/2$, this can be actually avoided by
requiring instead differentiability of \eq{cmap} as a map from
$\psi^2\phi
\Lambda^{2,k+2,\alpha}_{\phi,\psi,\varphi}(g_0)\times\psi^2\phi^2
\Lambda^{2,k+2,\alpha}_{\phi,\psi,\varphi}(g_0) $ to  $\psi^2
\Lambda^{1,k+1,\alpha}_{\phi,\psi,\varphi}(g_0)\times \psi^2
\Lambda^{0,k,\alpha}_{\phi,\psi,\varphi}(g_0)$.
\end{remark}
\begin{remark}
We note that a necessary condition for $\calSJr\ne \emptyset$ is
\bea &J_0\in \phi^{-2}C^{k+1,\alpha}_{\phi,1}(g_0)\;,
\quad\rho_0\in\phi^{-2} C^{k,\alpha}_{\phi,1}(g_0)\;.&\eeal{Jrho}
In any case it seems that the situation of main interest is
$J_0=\rho_0=0$.
\end{remark}

\begin{remark} The kernel in \eq{sub} is that of the operator
$P^*_{K_0+Q,g_0+h}$ acting from $H^{1}_{\phi,\psi}\times
H^{2}_{\phi,\psi}$  to $\phi^{-1}H^{0}_{\phi,\psi}\times\phi^{-2}
H^{0}_{\phi,\psi}$. We note that elliptic regularity shows that
elements of the kernel are as differentiable as the metric allows,
so the elements of the kernel are continuously differentiable
solutions  satisfying appropriate asymptotic properties.
\end{remark}
\begin{remark} We do not assume that
$(J_0,\rho_0)=(J,\rho)(K_0,\rho_0)$. Even if this last equality
holds,   $(0,0)$  will fail to be in $\cal S$ if there are KIDs at
$(K_0,g_0)$.
\end{remark}
\begin{remark}  We do not assume $(Q,h)$ to be small.
\end{remark}
\begin{remark}  Some rather general conditions which guarantee existence of
smoothing operators \eq{Psmooth} are given in
Appendix~\ref{Sconv}.
\end{remark}

\proof We wish to apply Lemma~\ref{LA1} with $x=(\delta K, \delta
g)$ and
$$u=Du(0)=\psi^2\Phi^2P^*_{\varepsilon,(K_0+Q_0,g_0+h_0)}\;,$$
$$v(x)=\psi^{-2}[(J,\rho)(K_0+Q_0+\delta K,g_0+h_0+\delta g)-(J,\rho)(K_0+Q_0,g_0+h_0)]\;,$$
$$Dv(0)=\psi^{-2}P_{K_0+Q_0,g_0+h_0}\;,$$
$$E=
\Lambda^{3,k+3,\alpha}_{\phi,\psi,\varphi}(g_0)\times
\Lambda^{4,k+4,\alpha}_{\phi,\psi,\varphi}(g_0)\;,$$
$$F=\psi^2\Phi
(\Lambda^{2,k+2,\alpha}_{\phi,\psi,\varphi}(g_0)\times
\Lambda^{2,k+2,\alpha}_{\phi,\psi,\varphi})(g_0)\;,$$
$$G=
\Lambda^{1,k+1,\alpha}_{\phi,\psi,\varphi}(g_0)\times
\Lambda^{0,k,\alpha}_{\phi,\psi,\varphi}(g_0)\;,$$
We start by verifying  that
$$L_\varepsilon:=\psi^{-2}P_{K_0+Q_0,g_0+h_0}\psi^2\Phi^2P^*_{\varepsilon,(K_0+Q_0,g_0+h_0)} $$
is an isomorphism from $E$ to $G$. We wish to use
Theorem~\ref{Til1projbis} with $(K_0,g_0)$ there replaced with
$(K_0+Q_0,g_0+h_0)$, and with $k$ there equal zero; the needed
regularity conditions on $\Ric(g_0+h_0)$ can be established by the
calculations of Lemma~\ref{samespace} using \eq{RicK} (recall that
$k\ge 2$), while the condition on $K_0+Q_0$ follows immediately
from \eq{RicK}. The remaining conditions are satisfied by
hypothesis. Since there are no KIDs we conclude that
$L_{\varepsilon=0}\equiv L_{\phi,\psi}$ (see \eq{uprightL}) is an
isomorphism  from
$\zH^{3}_{\phi,\psi}(g_0+h_0)\times\zH^{4}_{\phi,\psi}(g_0+h_0)$
to
$\zH^{1}_{\phi,\psi}(g_0+h_0)\times\zH^{0}_{\phi,\psi}(g_0+h_0)$;
those spaces coincide with the ones  based on $g_0$ by
Lemma~\ref{samespace}. The hypothesis of the existence of the
smoothing operation \eq{Psmooth} with $k$ there equal 0 implies
that $P_\varepsilon ^*\rightarrow_{\varepsilon \rightarrow0}P$ in
$$L(\zH^{3}_{\phi,\psi}\times\zH^{4}_{\phi,\psi},\phi^{-1}\zH^{2}_{\phi,\psi}\times\phi^{-2}\zH^{2}_{\phi,\psi})$$
implies that $L_\varepsilon$ is an isomorphism from
$\zH^{3}_{\phi,\psi}\times\zH^{4}_{\phi,\psi}$ to
$\zH^{1}_{\phi,\psi}\times\zH^{0}_{\phi,\psi}$ for $\varepsilon $
small enough. So $L_\varepsilon$ is injective on $E$. The weighted
elliptic regularity condition implies H\"older regularity of the
solution, and  surjectivity follows.

We leave it as an exercise to the reader to prove, using
Lemma~\ref{Lproduit} together with the arguments in
Lemma~\ref{samespace}, that:

\begin{Lemma}
\label{Lsmooth} Under  the conditions of
Theorem~\ref{submanifold}, the map \eq{cmap} is smooth from a
neighborhood of zero in $\psi^2\phi
C^{k+2,\alpha}_{\phi,\varphi}(g_0)\times\psi^2\phi^2
C^{k+2,\alpha}_{\phi,\varphi}(g_0) $ to  $\psi^2
C^{k+1,\alpha}_{\phi,\varphi}(g_0)\times \psi^2
C^{k,\alpha}_{\phi,\varphi}(g_0)$. \qed
\end{Lemma}

This does not suffice to prove differentiability of $v$, because
of the topology involved; however, differentiability with respect
to the Sobolev topology holds by hypothesis.

It follows that near $(Q_0,h_0)$ the set $\calSJr $ is an embedded
submanifold modeled on the kernel of $P_{K_0+Q_0,g_0+h_0}$. \qed

\begin{remark}\label{Rfol} The proof above actually gives a foliation of a neighborhood of
$(Q_0,h_0)$ in $F=\psi^2\Phi
(\Lambda^{2,k+2,\alpha}_{\phi,\psi,\varphi}(g_0)\times
\Lambda^{2,k+2,\alpha}_{\phi,\psi,\varphi}(g_0))\;$. Indeed, under
the conditions of Theorem~\ref{submanifold} we can use
Lemma~\ref{LA2} from Appendix~\ref{A} with the same spaces as
those  in Theorem~\ref{submanifold} and with $L=u$. For all
$(Q_0,h_0)\in{\cal S}_{(J_0,\rho_0)}$, there is a neighborhood $V$
 of zero in $G=
\Lambda^{1,k+1,\alpha}_{\phi,\psi,\varphi}(g_0)\times
\Lambda^{0,k,\alpha}_{\phi,\psi,\varphi}(g_0)\;,$ such that $$
\begin{array}{c}\left\{(Q,h)\in
\psi^2\phi \Lambda^{2,k+2,\alpha}_{\phi,\psi,\varphi}(g_0)\times
{\cal A}
^{k+2,\alpha}_{\phi,\psi,\varphi}(g_0),\right.\hspace{8cm} \\
\left.(J,\rho)(K_0+Q,g_0+h)=(J_0,\rho_0)+(\delta J,\delta \rho) ,
\;{ \Ker }\;P^*_{K_0+Q,g_0+h}=\{0\}\right\}_{(\delta J,\delta
\rho)\in V}
\end{array}$$
is a foliation. As we can do that for all points
$(Q_0,h_0)\in{\cal S}_{(J_0,\rho_0)}$, we obtain a foliation of a
neighborhood of ${\cal S}_{(J_0,\rho_0)}$ in $F$.

In fact, if we denote by $F_0$ the open subset of $\psi^2\phi
\Lambda^{2,k+2,\alpha}_{\phi,\psi,\varphi}(g_0)\times {\cal A}
^{k+2,\alpha}_{\phi,\psi,\varphi}(g_0)$ of elements $(Q,h)$ such
that
$$
\ker P^*_{K_0+Q,g_0+h}=\{0\}\;,
$$
then the map from $F_0$ to  $G$ defined by:
$$
f(Q,h)=(J,\rho)(K_0+Q,g_0+h)-(J,\rho)(K_0,g_0),
$$
is a submersion. In particular the levels sets
$$
\{(Q,h)\in F_0\;,\;\; f(Q,h)=(\delta J,\delta \rho)\}_{(\delta
J,\delta \rho)\in G},$$ provide a foliation of $F_0$.
\end{remark}

For completeness we note the following result:

\begin{Corollary}
\label{Csubmanifold}Every non-empty connected component of
$\calSJr$ defined in \eq{sub} is a Banach manifold modeled on the
kernel of $P_{K_0+Q_0,g_0+h_0}$ in
$\psi^2\Phi(\Lambda^{2,k+2,\alpha}_{\phi,\psi,\varphi}\times
\Lambda^{2,k+2,\alpha}_{\phi,\psi,\varphi})$ for an (arbitrarily
chosen) $(Q_0,h_0)\in \calSJr $.
\end{Corollary}

\proof An embedded submanifold of a Hausdorff space is necessarily
Hausdorff. The local coordinate patches are provided by the maps
which model $\cal S$ on its tangent spaces constructed in the
proof of  Theorem~\ref{submanifold}. It remains to show that all
the kernels are $(Q,h)$-independent. The proof of
Theorem~\ref{submanifold} shows that for all $(Q,h)\in \calSJr $
there exists a neighborhood thereof in $\calSJr $ which is
arc-connected. It then follows that a connected component of
$\calSJr $ is in fact arc-connected, by observing that the set of
metrics in $\calSJr $ which can be connected to a fixed metric in
$\calSJr $ by a continuous curve contained in $\calSJr $ is open
and closed in $\calSJr $. Thus for two couples $(Q_0,h_0)\in
{\calSJr}$ and $(Q_p,h_p)\in \calSJr $, there exists an arc
$\gamma$ in $\calSJr $ from one to the other. For each point
$x=(Q,h)\in\gamma$, there exists an $r_x>0$ such that
${\calSJr}\cap B_F(x,r_x)$  is diffeomorphic to an open subset of
the kernel of $P_{K+Q,g+h}$. As $\gamma$ is compact, there exists
a finite number of points $\{x_i=(Q_i,h_i)\in\gamma$, $i=0,..,p$\}
such that the union of the $B_F(x_i,r_{x_i})$'s covers $\gamma$.
This provides a finite chain of diffeomorphisms, a composition of
which identifies the kernel of $P_{(K+{Q_0},g+{h_0})}$ with that
of $P_{(K+{Q_p},g+{h_p})}$.

\qed

\section{Applications}\label{Sappli}

\subsection{Compact manifolds without boundary}
In this section  we apply Theorem~\ref{submanifold} to the case
where $M$ is a compact manifold without boundary. As already
pointed out, in this case we take
$$
\phi=\psi=\varphi=1,
$$
then the spaces we work with are the standard (non-weighted)
Sobolev and H\"older spaces.
\begin{theor}\label{Tcompact}
Let $k\geq 2$ and  $\alpha\in(0,1)$. Let $g_0\in C^{k+2,\alpha}$,
$K_0\in C^{k+2,\alpha}$. Then any non-empty connected component of
the set
$$
\begin{array}{c}\calSJr =\left\{(Q,h)\in
C^{k+2,\alpha}\times {\cal A}^{k+2,\alpha}_{1,1,1},\right.\hspace{5cm} \\
\mbox{ }\hspace{2cm}\left.(J,\rho)(K_0+Q,g_0+h)=(J_0,\rho_0), \;{
\Ker }\;P^*_{K_0+Q,g_0+h}=\{0\}\right\}\end{array}$$  is an
embedded submanifold of  $C^{k+2,\alpha}\times C^{k+2,\alpha}$.
\end{theor}

In Theorem~\ref{Tcompact} the kernel of the operator
$P^*_{K_0+Q,g_0+h}$ can be taken as that of an operator from
$H^{1}\times H^{2}$ to $H^{0}\times H^{0}$, or from $C^1\times
C^2$ to $C^0\times C^0$, or from $C^{k+1,\alpha}\times
C^{k+2,\alpha}$ to $C^{k,\alpha}\times C^{k,\alpha}$.

\subsection{Asymptotically flat manifolds without boundary}
\label{Saf} In this section  we apply Theorem~\ref{submanifold} to
the case where $M$ is an asymptotically flat manifold without
boundary and with compact interior; by definition, this means that
$M$ is the union of a compact set with a finite number of regions,
called ends, which are diffeomorphic to $\R^n\setminus B(0,R)$ for
some $R$. We denote by  $\hat \delta$  some arbitrarily chosen
 metric  of $C^{m+4,\alpha}$ differentiability class which
coincides with the Euclidean one in the asymptotically flat
regions. Here and in the following sections the index $m$
corresponds  to the differentiability of the background, typically
$m=\infty$ will be appropriate, however in some situations it
might be useful to have a background with finite
differentiability. The weight functions are $\phi=r$, with
$\varphi $ and $\psi$
--- powers of $r$ in the asymptotic regions, which are easily seen
to satisfy \eq{threecB}-\eq{scalprop3} in the asymptotic regions.
We extend the function $r$ to a smooth strictly positive function
in the compact region, then the requirements on the weight
functions are satisfied globally. It is convenient to relabel the
$\zHk _{r,r^{\alpha}}$ and the $C^{k,\alpha}_{r,r^{s}}$ spaces as
follows: choose some $m\in \N$, for $k\in \{0,...,m+4\}$,
$\alpha\in(0,1)$ and $\beta\in \R$ we set
$$ \zmcH _k^\beta = \zHk _{r,r^{-n/2-\beta}}(\hat \delta)\;,\;\;
C^{\beta}_{k,\alpha}=C^{k,\alpha}_{r,r^{-\beta}}(\hat \delta)\;.$$
For $\lambda\in\R$ we define \bel{Lambdef}
 \Lambda^{\lambda}_{k,\alpha}=\Lambda^{2,k,\alpha}_{r,r^{-n/2-\lambda},r^{-\lambda}}=\zmcH
_2^{\lambda}\cap C^{\lambda}_{k,\alpha}\;.\ee We also define for
$k\geq 2$:
\bel{bigM1}
\begin{array}{l}
{\cal M}_{k,\alpha}=\left\{ g \mbox{ is a metric uniformly
equivalent
to } \hat{\delta},\right.\\
\hspace{3cm}\left.\;g-\hat{\delta}\in
C^{0}_{k,\alpha},\;|\nabla^{(l)}_{\hat{\delta}}
(g-\hat{\delta})|_{\hat{\delta}}=o(r^{-l}),\; 0\leq
l\leq2\right\}\;. \end{array}\ee
 For $\gamma\geq0$ and
$k\leq m+2$ we let
$${\cal
A}_{k+2,\alpha}^{-\gamma}={\cal
A}^{2,k+2,\alpha}_{r,r^{n/2-2-\gamma},r^{-\gamma-2+n}}(\hat{\delta})\subset
C^{-\gamma}_{k+2,\alpha}\subset
 C^{0}_{k+2,\alpha}.$$
This corresponds to $\psi=r^{n/2-2-\gamma}$,
$\varphi=r^{-\gamma-2+n}$, and since $\phi = r$ the condition
\eq{threeaB} holds precisely for $\gamma \ge 0$. The choice of
weights here is justified by Theorem~7.7 in \cite{ChDelay}.

 With the labeling above, a metric such that $g-\hat \delta\in {\cal A}
_{k+2,\alpha}^{\beta}$ with $\beta\le 0$ clearly differs from the
Euclidean metric by $O(r^{\beta})$. In order to see that this is
actually $o(r^{\beta})$, for $i\in \N$ set
$$\Gamma_i\equiv \Gamma(2^iR,2^{i+1}R):=\overline{B(0,2^{i+1}R)}\backslash
\overline{B(0,2^iR)}\;.$$ Let $f\in \zmcH^\beta_0\cap C^\beta_1$
and let $x_n\in \overline{\Gamma}_n$ be any point such that
$$ |f(x_n)|= \sup_{y\in \Gamma_n} |f(y)|\;.$$
If $\|f\|_{C^\beta_1}=0$ there is nothing to prove, otherwise let
$$r_n = \min \left(\frac 18, \frac
{|f(x_n)|n^{-\beta}}{2\|f\|_{C^\beta_1}}\right)n\;.$$ For $y \in
B(x_n,r_n)$ we have
$$|f(y)-f(x_n)| \le \Big(\sup _{z \in  B(x_n,r_n)} |D f|(z)\Big)|y-x_n| \le C\|f\|_{C^\beta_1}n^{\beta-1} |y-x_n| \le
\frac{|f(x_n)|}2\;,$$ which implies
$$\int_{B(x_n,r_n)} f^2(y) |y|^{-2\beta-n}d^n y \ge
C'\left(\frac{f(x_n)}{2n^\beta}\right)^2\left(\frac{r_n}{n}\right)^n\;.
$$
The left-hand-side goes to zero as $n$ goes to infinity by the
dominated convergence theorem, which easily implies the result.

 The reference metric $g_0$ will be taken to be such
 that $g_0 \in {\cal M}_{k+2,\alpha}$.
 The
reference $K_0$ can be taken to be zero, but any $K_0\in
C_{k+2,\alpha}^{-1}$ with $K_0=o(r^{-1})$ will do; this last
condition ensures that the $\mcK$-set condition of
Theorem~\ref{submanifold} holds, see~\cite[Section~7]{ChDelay} for
details.

We take the smoothing operation to be the one in
Appendix~\ref{Sufc}, the only thing which needs to be checked is
the uniform covering condition \eq{uniform}: We write
$$\R^n\backslash
\overline{B(0,R)}=\cup_{i=0}^\infty \Gamma(2^iR,2^{i+1}R)\;.
$$
 Now $\Gamma(1,2)$ can be covered by a
finite number $N$ of ball's $B(x_j,|x_j|/8)$ with $x_j\in
\Gamma(1,2)$. Then $\Gamma(2^iR,2^{i+1}R)$ can be covered by $N$
balls $B(2^iR x_j,2^{i-3}R|x_j|)$ with $2^iR x_j\in
\Gamma(2^iR,2^{i+1}R)$. It is then clear that $\R^n\backslash
\overline{B}(0,R)$ can be covered by a countable set of balls
$B(y_k,|y_k|/8)$ with the property that for all $k\in\N$,
$$
\# \{l\;,\;\;B(y_l,|y_l|/2)\cap B(y_k,|y_k|/2)\neq\emptyset\}\leq
3N\;,
$$
as desired.


We note that the differentiability of the map \eq{cmap} follows
from the weighted equivalent of the Schauder ring property of
$H^k\cap L^\infty$. All the remaining hypotheses of
Theorem~\ref{submanifold} will be satisfied by~\cite{ChDelay}
(compare Section~7 there) under the following conditions:

\begin{theor}
\label{Taf} Let $m\in \N$, $k\in\{2,...,m\}$, $\alpha\in(0,1)$,
$g_0 \in {\cal M}_{k+2,\alpha}$, $K_0\in C_{k+2,\alpha}^{-1}$ with
$K_0=o(r^{-1})$. Let $\beta\ge 0$, $\beta\not\in\{n-2,n-1\}$. Then
any non-empty connected component of the set of KID-free initial
data:$$
\begin{array}{c}\calSJr =\left\{(Q,h)\in
\Lambda_{k+2,\alpha}^{-\beta-1}\times {\cal A}
_{k+2,\alpha}^{-\beta},\right.\hspace{5cm} \\
\mbox{ }\hspace{2cm}\left.(J,\rho)(K_0+Q,g_0+h)=(J_0,\rho_0), \;{
\Ker }\;P^*_{K_0+Q,g_0+h}=\{0\}\right\}\end{array}$$  is an
embedded submanifold of $\Lambda_{k+2,\alpha}^{-\beta-1}\times
{\cal A} _{k+2,\alpha}^{-\beta}$.

\qed
\end{theor}

In the above the kernel of the operator $P^*_{K_0+Q,g_0+h}$ is
viewed as that of an operator from ${\zmcH}_{1}^{\beta+2-n}\times
\zmcH_{2}^{\beta+2-n}$ to $\zmcH_{0}^{\beta-n+1}\times
\zmcH_{0}^{\beta-n}$. Elliptic regularity implies that elements of
this kernel are classically differentiable KIDs such that
$Y=o(r^{\beta+2-n})$, $N=o(r^{\beta+2-n})$. It is known that for
$0<\beta<n-2$  there are no non-trivial such KIDs, so we obtain:
\begin{cor}\label{coraf}
Under the conditions  of Theorem~\ref{Taf}, if $\beta\in (0,n-2)$
then all the level sets of the constraints map
$$
\left\{(Q,h)\in \Lambda_{k+2,\alpha}^{-t+1}\times {\cal A}
_{k+2,\alpha}^{-t+2},
(J,\rho)(K_0+Q,g_0+h)=(J_0,\rho_0)\right\}
$$
are embedded submanifolds of $ \Lambda_{k+2,\alpha}^{-t+1}\times
{\cal A} _{k+2,\alpha}^{-t+2}$.
\end{cor}

Thus, within the above set of weights the set of solutions of the
vacuum constraint equations does not have any manifold
singularities. On the other hand, such singularities will occur at
solutions with KIDs if higher values of $\beta$ are used. The
interest of such higher $\beta$'s relies in the fact that the
resulting manifolds of solutions possess fixed energy-momentum, or
angular momentum, or higher multipoles, depending upon the value
of $\beta$.

\subsection{ Compact manifold with boundary}\label{Scwb}
In this section  we apply Theorem~\ref{submanifold} to the case
where ${\overline M}$ is a compact manifold with smooth boundary
$\partial M$. We wish to construct manifolds of initial data on
$\overline M$ with prescribed boundary values on $\partial M$, as
well as a  prescribed number of transverse derivatives at the
boundary. Let $\gamma$ be any fixed auxiliary Riemannian metric of
$C^{m+4,\alpha}(\overline M)$ differentiability class. Let $x\ge
0$ be a function that vanishes precisely on $\partial M$, with
$dx$ nowhere vanishing there. We start by considering power-law
weighted
 spaces defined as
$$ \hbord^{s}_k=\zH^k_{1,x^{-s-n/2}}(\gamma)\;, \quad
 \cbord^s_{k,\alpha}=C^{k,\alpha}_{x,x^{-s}}(\gamma)\;,$$
 $$
 \Lambda^{s}_{k,\alpha}=\Lambda
^{2,k,\alpha}_{x,x^{-s-n/2},x^{-s}}=\hbord^{s}_2\cap
\cbord^{s}_{k,\alpha}\;.$$ We also define for  $m+4\geq k\geq 2$:
\bel{bigM2}
\begin{array}{l}
{\cal M}_{k,\alpha}=\left\{ g \mbox{ metric uniformly equivalent
to } \gamma,\right.\\
\hspace{3cm}\left.\;g-\gamma\in
\cbord^{0}_{k,\alpha},\;|\nabla^{(l)}_{\gamma}
(g-\gamma)|_{\gamma}=o(x^{-l}),\; 0\leq l\leq2\right\}.
\end{array}\ee
(this differs from \eq{bigM1} by a different background metric and
different functional spaces, we hope that this ambiguity will not
lead to confusions.)
 For $\sigma\geq 0$ and $m+2\geq k\geq 0$ we set
 $$
{\cal A} _{k+2,\alpha}^{\sigma}={\cal A}
^{k+2,\alpha}_{x,x^{\sigma-2+n/2},x^{\sigma-2+n}}(\gamma)\subset
\cbord^{\sigma}_{k+2,\alpha}\subset \cbord^0_{k+2,\alpha}\;.$$

The first inclusion shows that metrics of the form $g=\gamma+h$
with $ h\in {\cal A} _{k+2,\alpha}^{\sigma}$ approach  $\gamma$ at
$\partial M$ at least as $O(x^\sigma)$, and in fact an argument
similar to the one in Section~\ref{Saf} shows that this is
actually at least $o(x^\sigma)$. The above corresponds to
$\phi=x$, $\varphi=x^{\sigma+n-2}$, $\psi=x^{\sigma+n/2-2}$, with
the choices being justified as follows:
In~\cite[Theorem~5.6]{ChDelay} we obtain metrics such that $\delta
g$  is in $\zH_{x,x^{-(s-n+2)-n/2}}\cap C_{x,x^{-(s-n+2)}}$. It is
convenient to number the spaces according to the decay rate of
$\delta g$ near the boundary, so we set
 $\sigma=s-n+2$.
 In the current set-up we have $\delta g$  in
$\psi^2\phi^2(\zH_{\phi,\psi}\cap C_{\phi,\varphi})$, which after
straightforward algebra uniquely leads to the weights above.

In order to obtain the required smoothing operator we use the
results in Appendix~\ref{Sufc}, we need to justify the covering
condition there. Let $T>0$, we set
$$ \R^n_+=\{x=(x^1,...,x^n)\}\in\R^n, x^n>0\}=
\cup_{i=-\infty}^\infty{\cal B}(2^{i}T,2^{1+i}T),$$ where ${\cal
B}(2^{i}T,2^{i+1}T)=\{x=(x^1,...,x^n)\in\R^n, 2^{i}T <x^n\leq
2^{i+1}T\}$. We have  $\partial \R^n_+=\{x\in\R^n, \;  x^n=0\}$.
${\cal B}(1,2)$ can be covered by closed cubes with edge sizes
one, with pairwise intersections empty, or along the faces of the
cubes. We choose one of those cubes, call it $K_1$, and we cover
it by $N(n)$ balls $B(x_i,|x_i|/8)$ with $x_i\in K_1$; every other
cube in ${\cal B}(1,2)$ will then be covered by $N(n)$ balls by
translating the balls covering $K_1$. We cover ${\cal B}(1/2,1)$
with  cubes of edge sizes $1/2$, intersecting along the faces
only, such that a cube of ${\cal B}(1,2)$ intersects precisely
$2^{n-1}$ cubes of ${\cal B}(1/2,1)$. Each cube of ${\cal
B}(1/2,1)$ can be covered by $N(n)$ balls $B(x_i,|x_i|/8)$ with
$x_i$ in that cube  by scaling by $1/2$ and translating the balls
covering $K_1$. An inductive repetition of the procedure leads to
a covering of $\R^n_+$ by a countable set of balls
$B(y_k,|y_k|/8)$ with the property that for all $k$,
$$ \# \{l\;,\;\;B(y_l,|y_l|/2)\cap
B(y_k,|y_k|/2)\neq\emptyset\}\leq C_nN(n),$$ for some constant
$C_n$. Working in local charts, and using partitions of unity, the
above construction provides the required covering near the
boundary of a manifold.

We refer the reader to \cite[Section~5]{ChDelay} for a
justification of the remaining hypotheses of
Theorem~\ref{submanifold}:

\begin{theor}\label{T6.4}
Let  $k\in\{2,...,m\}$,  $\alpha\in(0,1)$, $g_0\in {\cal M}
_{k+2,\alpha}$, $K_0 \in \cbord_{k+2,\alpha}^{-1}$, with {\rm
\beaa 
& x|K_0|_{g_0}+x^2|\nabla K_0|_{g_0}\to_{x\to 0}0\;.&
\eeaa}For $\sigma\geq0$ when $n>3$ or $\sigma>0$  any non-empty
connected component of
$$
\begin{array}{c}\calSJr =\left\{(Q,h)\in
\Lambda_{k+2,\alpha}^{\sigma-1}\times {\cal A}
_{k+2,\alpha}^{\sigma},\right.\hspace{5cm} \\
\mbox{ }\hspace{2cm}\left.(J,\rho)(K_0+Q,g_0+h)=(J_0,\rho_0), \;{
\Ker }\;P^*_{K_0+Q,g_0+h}=\{0\}\right\}\end{array}$$ is an
embedded submanifold of $ \Lambda_{k+2,\alpha}^{\sigma-1}\times
{\cal A} _{k+2,\alpha}^{\sigma}$.
\end{theor}
%

Here the kernel of the operator $P^*_{K_0+Q,g_0+h}$ is, by
elliptic regularity, a subspace of
$\left({\hbord}_{1}^{-\sigma-n+2}\times
\hbord_{2}^{-\sigma-n+2}\right)\cap
\left(C^{k+2}(\overline{M})\times C^{k+2}(\overline{M})\right)$.

For further reference we note the following result:

 \begin{Proposition}
 \label{Psio} Suppose that $(K_0,g_0)\in \left(C^{k+2,\alpha}\times
 C^{k+2,\alpha}\right)(M)$, $k\ge 2$, $\alpha \in (0,1)$, and let $\Omega\subset M$ be a domain with
 smooth boundary and compact closure.
For all $s\neq (n+1)/2,(n+3)/2$, the image of the linearisation
$P$, at $(K_0,g_0)$, of the constraints map, when defined on $
\left(\Lambda^{-s+1}_{k+2,\alpha}\times
\Lambda^{-s+2}_{k+2,\alpha}\right)(\Omega)$, is
$$\left(x^{n-2s}{\maclKzoz }\right)\cap (\Lambda^{-s}_{k+1,\alpha}\times
\Lambda^{-s}_{k,\alpha})\;.$$  Here ${\cal K}_0$ is the space of
KIDs which are in $\hbord^{s-n}_{1}\times \hbord^{s-n}_{2}\subset
\left(L^2\times L^2\right)(\Omega,x^{-2s+n}d\mu_{g_0})$, and
orthogonality is taken in $\left(L^2\times
L^2\right)(\Omega,x^{-2s+n}d\mu_{g_0})$. In other words, the image
of $P$ is$$
\begin{array}{c}\left\{(J,\rho)\in \Lambda^{-s}_{k+1,\alpha}\times
\Lambda^{-s}_{k,\alpha}\ \mbox{ such that } \
\langle(J,\rho),(Y,N)\rangle_{(L^2\oplus L^2)(\Omega,d\mu_{g_0})}=0\;\right.\\
\left. \mbox{ for all } \  (Y,N)\in H^{s-n}_1\times H^{s-n}_2 \
\mbox{ satisfying } \  P^*(Y,N)=0\right\}\;. \end{array}$$ Further
$P^{-1}(0)\subset \Lambda^{-s+1}_{k+2,\alpha}\times
\Lambda^{-s+2}_{k+2,\alpha}$ splits.
 \end{Proposition}

 \proof The proof of this
result is essentially contained in that of Theorem~\ref{T6.4}; the restriction 
on the constant $\sigma$ there arises from the requirement that
the full non-linear constraint map be well defined, but this
restriction is  not needed for the linearised problem.  We note
that a closed space complementing $P^{-1}(0)$ is provided by
$\mbox{\rm Im}\, (\Phi^2\psi^2 P^*_\epsilon)$, for $\epsilon$
small enough (compare the arguments in Appendix~\ref{A}), and that
$P$ restricted to this space is an isomorphism onto Im$(P)$.\qed

Alternative useful weights are the exponential ones, for those we
will use  Proposition \ref{regulconvbis} to obtain the needed
regularisation. We take the following weight
functions\footnote{Theorem 5.9 and Proposition 5.10 of
\cite{ChDelay} provide $\delta g$ in
$\psi^2\phi^2(\zH_{\phi,\psi}\cap C_{\phi,\varphi})=x^4
(\zH_{x^2,x^{-n}e^{s/x}}\cap C_{x^2,e^{s/x}})$, which leads to the
choices of the weights above.}
$$
\phi=x^2\;,\;\;\psi=x^{n}e^{-s/x}\;,\;\;\varphi=x^{2n}e^{-s/x}.$$
With those choices we note that
$$ {\cal A}
^{k+2,\alpha}_{x^2,x^{n}e^{-s/x},x^{2n}e^{-s/x}}\subset
\psi^2\phi^2(H_{\phi,\psi}\cap C_{\phi,\varphi})=
\phi^2(H_{\phi,\psi^{-1}}\cap C_{\phi,\varphi\psi^{-2}})=
x^4(\zH^{2}_{x^2,x^{-n}e^{s/x}}\cap C^{k+2,\alpha}_{x^2,e^{s/x}}).
$$
We define the space
$$
\begin{array}{l}
{\cal M}^{exp}_{k,\alpha}=\left\{ g \mbox{ metric uniformly
equivalent
to } \gamma,\right.\\
\hspace{4cm}\left.\;g-\gamma\in
C^{k,\alpha}_{x^2,1},\;|\nabla^{(l)}_{\gamma}
(g-\gamma)|_{\gamma}=o(x^{-2l}),\; 0\leq l\leq2\right\}.
\end{array}
$$
Using the results in Section~5 of~\cite{ChDelay} one now has:
\begin{theor} Let $k\in\{2,...,m\}$, $\alpha\in(0,1)$, $g_0\in
{\cal M}^{exp}_{k+2,\alpha}$, $K_0 \in
x^{-2}C^{k+2,\alpha}_{x^2,1}$, with $$
x^2|K_0|_{g_0}+x^4|\nabla K_0|_{g_0}\to_{x\to 0}0\;. $$ 
For $s>0$ any non-empty connected component of
$$
\begin{array}{c}\calSJr =\left\{(Q,h)\in
x^2\Lambda^{2,k+2,\alpha}_{x^2,x^{-n}e^{s/x},e^{s/x}}\times {\cal
A}
^{k+2,\alpha}_{x^2,x^{n}e^{-s/x},x^{2n}e^{-s/x}},\right.\hspace{5cm} \\
\mbox{ }\hspace{2cm}\left.(J,\rho)(K_0+Q,g_0+h)=(J_0,\rho_0), \;{
\Ker }\;P^*_{K_0+Q,g_0+h}=\{0\}\right\}\end{array}$$ is an
embedded submanifold of
$x^2\Lambda^{2,k+2,\alpha}_{x^2,x^{-n}e^{s/x},e^{s/x}}\times {\cal
A} ^{k+2,\alpha}_{x^2,x^{n}e^{-s/x},x^{2n}e^{-s/x}}$.
\end{theor}
Here the kernel of the operator $P^*_{K_0+Q,g_0+h}$ is a subspace
of $\zH^{1}_{x^2,x^{n}e^{-s/x}}\times \zH^2 _{x^2,x^{n}e^{-s/x}}$;
but elliptic regularity shows that elements of the kernel are
classically differentiable in the interior, and it is standard to
show that they are in $C^{k+2}(\overline{M})\times
C^{k+2}(\overline{M})$.

\subsection{ Conformally compactifiable manifolds}
In this section  we apply Theorem~\ref{submanifold} to the case
where $M$ is a conformally compactifiable manifold (with a compact
conformal boundary at infinity), as in Section~6 of
\cite{ChDelay}. Let $\gamma$ be $ C^{m+4,\alpha}$ conformally
compactifiable, so that $\gamma = x^{-2}\overline \gamma$, with
$x$ -- defining function for the conformal boundary $\partial M$,
and $\overline \gamma$ -- a Riemannian metric on $\overline M$ of
$ C^{m+4,\alpha}(\overline M)$ differentiability class.
 In that context, it is  natural  to
define $$
\hhyp^s_k:=\zH^k_{1,x^{-s}}(\gamma)=\hbord^{s}_k(\overline{\gamma})
\;,
$$
 $$
 \chyp^s_{k,\alpha}:=C^{k,\alpha}_{x,x^{-s}}(\overline{\gamma})=\cbord^s_{k,\alpha}(\overline{\gamma}),$$
 $$\Lambda _{k,\alpha}^{s}=\Lambda
^{2,k+2,\alpha}_{1,x^{-s},x^{-s}}(\gamma)=\chyp^s_{k,\alpha}\cap\hhyp^s_2,$$
 and for $t\geq0$,
 $$
{\cal A} _{k+2,\alpha}^{t}={\cal A}
^{k+2,\alpha}_{1,x^{t},x^t}(\gamma)\subset\chyp^t_{k+2,\alpha}\subset
\chyp^0_{k+2,\alpha} .$$ The space ${\cal M}_{k,\alpha}$ is
defined similarly to \eq{bigM2}, but both the background $\gamma$
and the function spaces involved are different now:
$$
\begin{array}{l}
{\cal M}_{k,\alpha}=\left\{ g \mbox{ metric uniformly equivalent
to } \gamma,\right.\\
\hspace{4cm}\left.\;g-\gamma\in
\chyp^0_{k,\alpha},\;|\nabla^{(l)}_{\gamma}
(g-\gamma)|_{\gamma}=o(1),\; 0\leq  l\leq2\right\}. \end{array}
$$
The same locally uniform covering  as in the preceding section can
be used here, so the smoothing operator of Appendix~\ref{Sufc}
applies, leading to:

\begin{theor} Let $k\in \{2,...,m\}$ and $\alpha\in(0,1)$. Let
$g_0\in{\cal M}_{k,\alpha}$, $K_0=\lambda_0g_0+L_0$ with
$L_0,\lambda_0 \in \chyp_{k+2,\alpha}^{0}$,
$|L_0|_{\gamma}\rightarrow_{x\rightarrow0}0$, $|\nabla
L_0|_{\gamma}\rightarrow_{x\rightarrow0}0$.  Let $t\geq0$,
$t\not\in\{(n-3)/2,(n-1)/2,(n+1)/2\}$. Then any connected
component of the set $$
\begin{array}{c}\calSJr =\left\{(Q,h)\in
\Lambda_{k+2,\alpha}^{t}\times {\cal A}
_{k+2,\alpha}^{t},\right.\hspace{5cm} \\
\mbox{ }\hspace{2cm}\left.(J,\rho)(K_0+Q,g_0+h)=(J_0,\rho_0), \;{
\Ker }\;P^*_{K_0+Q,g_0+h}=\{0\}\right\}\end{array}$$ is a
submanifold of $ \Lambda_{k+2,\alpha}^{t}\times {\cal A}
_{k+2,\alpha}^{t}$. For $0\le t < (n+1)/2$ the kernel condition is
automatically satisfied.
\end{theor}
Here the kernel of  $P^*_{K_0+Q,g_0+h}$ is that for a map from
${\hhyp}_{1}^{-t}\times \hhyp_{2}^{-t}$  to $\hhyp_{0}^{-t}\times
\hhyp_{0}^{-t}$; this can also be reformulated in terms of
classical differentiability in appropriately weighted spaces.

Remarks similar to those following Corollary~\ref{coraf}
concerning the value of $t$ apply here.

\begin{remark}
For $(Q,h)\in{\cal S}_{(J_0,\rho_0)}$ we have that
$|Q|_\gamma=o(1)$ and   $|\nabla Q|_\gamma=o(1)$.
\end{remark}

\appendix

\section{Submanifolds, foliations}\label{A}

The following is a variation of an argument
in~\cite{MR2000j:53043}:
\begin{lem}\label{LA1}
Let $E$, $F$ and $G$ be three Banach spaces, and let $u$ (resp.
$v$) be a map defined from a neighborhood of\/ $0$ in $E$ (resp.
$F$) to $F$ (resp. $G$) such that $u(0)=0$ (resp. $v(0)=0$) and
which is differentiable at $0$. We also assume  that $Dv(0)\circ
Du(0)$ is an isomorphism from $E$ to $G$. Then the set $v^{-1}(0)$
is a submanifold of $F$ in a neighborhood of $0$.
\end{lem}
\proof For $x\in F$,
$$
x=\underbrace{Du(0)\circ[Dv(0)\circ Du(0)]^{-1}\circ
Dv(0)(x)}_{\in \; \Ima\;Du(0)}+\underbrace{x-Du(0)\circ[Dv(0)\circ
Du(0)]^{-1}\circ Dv(0)(x)}_{\in \;\Ker\;Dv(0)}\;.$$ (It
easily follows that $F=\Ima\;Du(0)\oplus \Ker Dv(0)$, with both
summands closed.) As $Dv(0)\circ Du(0)$ is an isomorphism, the
inverse function theorem shows that $v\circ u$ is a diffeomorphism
in a neighborhood of $0$, so for $x\in F$ close to zero we have
$$
x=\underbrace{u\circ[v\circ u]^{-1}\circ v(x)}_{\in
\;\Ima\;u}+{x-u\circ[v\circ u]^{-1}\circ v(x)}\;.$$ Let us define
a map from a neighborhood of zero in $F$ to $F$ by
$$
f(x)=x+u\circ[v\circ u]^{-1}\circ v(x)-Du(0)\circ[Dv(0)\circ
Du(0)]^{-1}\circ Dv(0)(x)\;.$$ One clearly has
$$
Df(0)=\mbox{\rm Id}\;,$$ and the inverse function theorem shows
that $f$ is a diffeomorphism in a neighborhood of zero. We also
have
$$
x\in v^{-1}(0)\iff f(x)\in Ker\; Dv(0)\;,
$$
(for the "$\Leftarrow$" part we use the fact that $Dv(0)\circ
u\circ[v\circ u]^{-1}$ is an isomorphism near zero), so $f$
provides the required map modeling  $v^{-1}(0)$ on a linear space.
\qed

Lemma~\ref{LA1} shows how to straighten-up a level set of $v$; one
can similarly show existence of foliations by level sets:

\begin{lem}\label{LA2} Let $E$, $F$ and $G$ three Banach spaces, $L$  a
linear continuous map  from  $E$ to $F$, $v$ a map defined from a
neighborhood of a point $x_0$ in  $F$ to $G$,  continuously
differentiable near $x_0$. We  assume  that $Dv(x_0)\circ L$ is an
isomorphism from $F$ to $G$. Then there exist a neighborhood ${V}$
of $y_0=v(x_0)$ in $G$ such that the collection of level-sets of
$v$, $$ \{x\in F\;,\;\; v(x)=y\}_{y\in {V}},$$ is a foliation of a
neighborhood $U$ of $x_0$  in $F$.
\end{lem}

 \proof We have that $Dv(x)\circ L$ is a diffeomorphism for
$x\in F$ close to $x_0$, one then has as in Lemma~\ref{LA1} that
$$
F=\Ima L\oplus \Ker Dv(x),
$$
and one easily checks that $\Ima L$ is closed (recall that all
maps are continuous). So $Dv(x)$ is surjective and its  kernel
splits. From~\cite[p.~21]{Lang} the map $v$ is then a submersion
near $x_0$. In particular, again from~\cite[p.~20]{Lang}, there
exist $U$ neighborhood of $x_0$ in $F$, $V$ neighborhood of $y_0$
in $G$ and two isomorphisms $\varphi: U\longrightarrow U_1\times
U_2$ ($U_1$ and $U_2$ open subset of some Banach spaces) and
$\psi:V\longrightarrow V_2$ ($V_2$ open subset of some Banach
spaces with $U_2\subset V_2$) such that
$$
\psi\circ v\circ\varphi^{-1}:U_1\times U_2\longrightarrow V_2,
$$
is the  projection on the second axis. This  gives the desired
foliation of $U$. \qed

\section{Two weighted smoothing operators}
\label{Sconv}

 In this appendix we will show how to define  smoothing operators as needed
 in the body of the paper; this will require a set of
conditions on the functions $\phi $ and $\varphi $, compatible
with the usual settings of interest in general relativity. The
technique of Appendix~\ref{Sacm} seems to be somewhat simpler than
that of Appendix~\ref{Sufc}, and does not require any covering
conditions. However, covering conditions arise naturally when
regularising functions in weighted Sobolev classes, therefore it
seemed of interest to us  to present both methods.

\subsection{Smoothing with locally uniformly finite coverings}
\label{Sufc}

Throughout this appendix we assume that the manifold $M$ is an
open subset of $\R^n$, equipped with an Euclidean metric (which is
of course not the physical space metric we are interested in), and
we will be regularising functions. The regularisation can then be
applied to tensor fields on more general manifolds by using
coordinate patches, partitions of unity, and usual covering
arguments.

We assume that $\phi $ and $\varphi $ verify
\eq{threecB}-\eq{scalprop2}. For all $p\in M$, we denote by $B_p
$, the open ball of center $p$ with radius $\phi(p) /2$. We
require that\footnote{\Eq{scalprop0} can be replaced by the weaker
condition that there exists $\mu>0$ such that for all $p\in M $ we
have $B(p,\mu\phi(p) ) \subset M$, as changing $\phi$ to $\mu
\phi$ for a positive constant $\mu$ leads to equivalent norms. So,
\emph{e.g.} in the asymptotically flat case, one actually has to
replace the weight $\phi=r$ for $r\ge R$ by $\phi=r/2R$. Any such
rescaling lead to obvious changes in the hypotheses needed for the
covering arguments below.} for all $p\in M $,
\bel{scalprop0}B(p,\phi(p) ) \subset M\;.\ee

For $\rho\in[1,\infty)$ we shall use the following notation:
$$
B_i^\rho=B(p_i,\phi(p_i)/\rho).$$ Our next restriction is that the
manifold can be covered by a countable collection of balls
$B^8_i$, \bel{union} M=\cup_{i=1}^\infty B^8_i\;,\ee such that
there exists an $N\in\N$ so that for all $i\in\N$, \bel{uniform}
\#\{j\;, B^2_i\cap B^2_j\neq\emptyset\}\leq N\;.\ee
For $p\in M $, we set
$$
\varphi_p:B(0,1/2)\ni z\mapsto p+\phi(p) z\in B_p \;.
$$
This implies that for all functions $u$ on $M$ and all
multi-indices $\gamma$ we have
$$
\partial_z^\gamma(u\circ
\varphi_p )=\phi(p) ^{|\gamma|}(\partial^\gamma u)\circ
\varphi_p\; .
$$
Using \eq{scalprop1} and \eq{scalprop2} it is easy to see that we
have:
\begin{lem}\label{equivnorm} For  $\rho\in[2,8]$ the following
norms on  $C^{k,\alpha}_{\phi ,\varphi }(M)$ are equivalent:
$$
\begin{array}{lll}
\|u\|_{C^{k,\alpha}_{\phi ,\varphi }(M)}&\sim&\sup_{i\in \N
}\|u\|_{C^{k,\alpha}_{\phi ,\varphi }(B^\rho_i)}\\
&\sim&\sup_{i\in \N }\|u\|_{C^{k,\alpha}_{\phi(p_i) ,\varphi(p_i)
}(B^\rho_i)}\\&\sim& \sup_{i\in \N
}\|u\circ\varphi_{p_i}\|_{C^{k,\alpha}_{1,\varphi(p_i)
}(B(0,1/\rho))}.
\end{array}
$$
\end{lem}
We now construct a convenient partition of unity:
\begin{lem}\label{partition}
There exists a partition of unity
$$
\sum_{i=1}^\infty \zeta_i=1,
$$
with smooth functions $\zeta_i\geq0$, and 
$\zeta_i=0$ outside $B^4_i$,  such that for all $l\in\N$ and
$\alpha\in(0,1)$ there exists a constant $C(l,\alpha)$ so that for
all $i\in\N$,
$$
\|\zeta_i\|_{C^{l,\alpha}_{\phi , 1 }(M)}\leq C(l,\alpha).
$$
\end{lem}
\proof Let $\chi$ be a smooth non-negative function on $\R^n$ such
that $\chi=1$ on $B(0,1/8)$ and $\chi=0$ outside $B(0,1/4)$. We
define
$$
\chi_i:=\chi\circ\varphi^{-1}_{p_i}\;,\quad
\zeta_i:=\frac{\chi_i}{\sum_{j=1}^\infty\chi_j}.
$$
Let us show that the sum in the definition above is well defined
and greater than one. If $x\in B^4_i$, there exists at most $N$
balls $B^4_j$, with $N$ given by \eq{uniform}, such that $x\in
B^4_j$. Since $\chi_j$ has support in $B^4_j$, the sum at $x$ is
over a finite set. If $x\in B^8_i$ then $\chi_i(x)=1$ thus the sum
is not less than one. If $x\in B^4_i\backslash B^8_i$, from
\eq{union}, $x$ must be in some $B^8_j$ thus $\chi_j(x)=1$ and
then, again, the sum is greater than or equal to one.

Now,  $\chi_i$ has compact support so is in $C^{l,\alpha}_{\phi ,
1 }(M)$, and from Lemma~\ref{equivnorm}
$$
\|\chi_i\|_{C^{l,\alpha}_{\phi , 1 }(M)}=
\|\chi_i\|_{C^{l,\alpha}_{ \phi, 1 }(B^4_i)}\leq C'
 \|\chi_i\circ\varphi_{p_i}\|_{C^{l,\alpha}_{ 1, 1 }(B(0,1/4))}=
 C'\|\chi\|_{C^{l,\alpha}_{ 1, 1 }(B(0,1/4))}=:C_1,
$$
where $C_1$ depends upon $l$ and $\alpha$ but does not depend upon
$i$. So we have that
$$
\|\sum_{j=1}^\infty\chi_j\|_{C^{l,\alpha}_{ \phi, 1 }(B^2_i)}\leq
 \sum_{\{j\;, B^2_i\cap B^2_j\neq\emptyset\}}\|\chi_j\|_{C^{l,\alpha}_{ \phi, 1
 }(M)}\leq NC_1.
$$
and thus
$$
\|\sum_{j=1}^\infty\chi_j\|_{C^{l,\alpha}_{ \phi, 1 }(M)}\leq
  NC_1.
$$
Finally, as the sum is greater than or equal to one it is easy to
see that the $\zeta_i$'s satisfy the desired properties.
\qed 

Let $\theta$ by any smooth strictly positive function on $\R^n$,
with support in $B(0,1)$ and such that
$$
\int_{\R^n}\theta=1.$$ For $\varepsilon>0$ we set $$
\theta_\varepsilon(x)=\frac{1}{\varepsilon^n}\theta(\frac{x}{\varepsilon})\;.$$
For $u$ a function on $M$ and $\varepsilon>0$, we define
$$
u_i=\zeta_i u\;,\;\;\;\hat{u}_i=u_i\circ\varphi_{p_i},$$
$$
\hat{u}_{i,\varepsilon}=\theta_\varepsilon*\hat{u}_i\;,\;\;
u_{i,\varepsilon}=\hat{u}_{i,\varepsilon}\circ\varphi_{p_i}^{-1}\;,\;\;
u_\varepsilon=\sum_{i=1}^\infty u_{i,\varepsilon}
$$
\begin{prop}\label{regulconv}
Let $u\in C^{k,\alpha}_{\phi ,\varphi }(M)$. For all
$\varepsilon\in(0,1/4)$ and all $m\in\N$ we have $u_\varepsilon\in
C^{m}_{\phi ,\varphi }(M)$. Further, $u_\varepsilon$  converges to
$u$ in $C^{k,\alpha}_{\phi ,\varphi }(M)$ as $\varepsilon$ goes to
zero.
\end{prop}
\proof First remark that as the $u_i$'s have support in $B^4_i$,
then for $\varepsilon<1/4$ the functions $u_{i,\varepsilon}$ have
support in $B^2_i$. It follows that  on $B^2_i$ we have
$$
u_\varepsilon=\sum_{\{j\;, B^2_i\cap
B^2_j\neq\emptyset\}}u_{j,\varepsilon}.$$ Lemma~\ref{equivnorm}
and \ref{partition} together with standard properties of
convolution in $\R^n$ imply
$$
\begin{array}{lll}
\|u_\varepsilon\|_{C^{k,\alpha}_{\phi ,\varphi }(B^2_i)}&\leq&
\sum_{\{j\;, B^2_i\cap
B^2_j\neq\emptyset\}}\|u_{j,\varepsilon}\|_{C^{k,\alpha}_{\phi
,\varphi
}(B^2_i)}\\
&=&\sum_{\{j\;, B^2_i\cap
B^2_j\neq\emptyset\}}\|u_{j,\varepsilon}\|_{C^{k,\alpha}_{\phi
,\varphi
}(B^2_i\cap B^2_j)}\\
&\leq &\sum_{\{j\;, B^2_i\cap
B^2_j\neq\emptyset\}}\|u_{j,\varepsilon}\|_{C^{k,\alpha}_{\phi
,\varphi
}(B^2_j)}\\
&\leq& C\sum_{\{j\;, B^2_i\cap
B^2_j\neq\emptyset\}}\|\hat{u}_{j,\varepsilon}\|_{C^{k,\alpha}_{\phi(p_j)
,\varphi(p_j) }(B(0,1/2))}\\
&\leq&C\sum_{\{j\;, B^2_i\cap
B^2_j\neq\emptyset\}}\|\hat{u}_{j}\|_{C^{k,\alpha}_{\phi(p_j)
,\varphi(p_j) }(B(0,1/2))}\\
&\leq&C'\sum_{\{j\;, B^2_i\cap
B^2_j\neq\emptyset\}}\|{u}_{j}\|_{C^{k,\alpha}_{\phi
,\varphi }(B^2_i)}\\
&\leq& N C''C(k,\alpha)\|u\|_{C^{k,\alpha}_{\phi ,\varphi }(M)}\;.
\end{array}
$$
In particular $u_\varepsilon\in C^{k,\alpha}_{\phi ,\varphi }(M)$,
with norm uniformly bounded in $\varepsilon$. Let us now show that
in fact $u_\varepsilon$ is also in $ C^{k+l}_{\phi ,\varphi }(M)$
for any $l\geq 0$. First, we have:
$$
\begin{array}{lll}
\sup_{B^2_i} |\varphi\phi^{k+l}\partial^{k+l}
u_{i,\varepsilon}|&\leq &C \sup_{B^2_i}
|\varphi(p_i)\phi^{k+l}(p_i)\partial^{k+l} {u}_{i,\varepsilon}|
\\&=& C \sup_{B(0,1/2)}
|\varphi(p_i)\partial^{k+l} \hat{u}_{i,\varepsilon}|\\
&\leq &C
\|\partial^l\theta_\varepsilon\|_{L^1(B(0,1/2))}\|\varphi(p_i)\partial^k\hat{u}_i\|_{L^{\infty}(B(0,1/2))}\\
&=&C\displaystyle{\frac{1}{\varepsilon^l}\|\partial^l\theta\|_{L^1(\R^n)}\|\varphi(p_i)\partial^k\hat{u}_i\|_{L^{\infty}(B(0,1/2))}
}\\
&\leq&
\displaystyle{C'\frac{C_l}{\varepsilon^l}\|\varphi\phi^k\partial^k
u_i\|_{L^{\infty}(B^2_i)}}\\&\leq&
C''\displaystyle{\frac{C_l}{\varepsilon^l}\|u\|_{C^{k}_{\phi,\varphi}(M)}}.
\end{array}
$$
Thus we have
$$
\sup_{B^2_i}|\varphi\phi^{k+l}\partial^{k+l} u_\varepsilon|\leq
\sum_{\{j\;,\;\;B^2_i\cap B^2_j\neq \emptyset\}} \sup_{B^2_i\cap
B^2_j }|\varphi\phi^{k+l}\partial^{k+l} u_{j,\varepsilon}|\leq N
C'''\displaystyle{\frac{C_l}{\varepsilon^l}\|u\|_{C^{k}_{\phi,\varphi}(M)}}\;,
$$
and then, using Lemma~\ref{equivnorm},
$$
 \sup_{M}|\varphi\phi^{k+l}\partial^{k+l}
u_\varepsilon|\leq N
C'''\displaystyle{\frac{C_l}{\varepsilon^l}\|u\|_{C^{k}_{\phi,\varphi}(M)}}\;,
$$
which gives $u_\varepsilon\in C^{k+l}_{\phi,\varphi}(M)$ for any
$l\geq0.$

Let us pass now to the proof that $u_\varepsilon$ converges to $u$
in $C^{k,\alpha}_{\phi,\varphi}(M)$. We have
$$
\begin{array}{lll}
\|u_{i,\varepsilon}-u_i\|_{C^{k,\alpha}_{\phi,\varphi}(B^2_i)}&\leq&
C
\|\hat{u}_{i,\varepsilon}-\hat{u}_i\|_{C^{k,\alpha}_{1,\varphi(p_i)}(B(0,1/2))}\\
&\leq&
C'\varepsilon^\alpha\|\hat{u}_i\|_{C^{k,\alpha}_{1,\varphi(p_i)}(B(0,1/2))}\\
&\leq&
C''\varepsilon^\alpha\|{u}\|_{C^{k,\alpha}_{\phi,\varphi}(M)}.\\
\end{array}
$$
Thus as $u_\varepsilon-u=\sum_{i=1}^\infty
(u_{i,\varepsilon}-u_i)$, on $B^2_i$, we have
$$
\|u_{\varepsilon}-u\|_{C^{k,\alpha}_{\phi,\varphi}(B^2_i)}\leq
\sum_{\{j\;,\;\;B^2_i\cap
B^2_j\neq0\}}\|u_{j,\varepsilon}-u_j\|_{C^{k,\alpha}_{\phi,\varphi}(B^2_i\cap
B^2_j )}\leq
NC''\varepsilon^\alpha\|{u}\|_{C^{k,\alpha}_{\phi,\varphi}(B^2_i)}\;,
$$
then finally
$$
\|u_{\varepsilon}-u\|_{C^{k,\alpha}_{\phi,\varphi}(M)}\leq
NC'\varepsilon^\alpha\|{u}\|_{C^{k,\alpha}_{\phi,\varphi}(M)}\;.
$$
\qed

\subsection{Smoothing using a weighted convolution}
\label{Sacm}

For the exponential weights considered in Section~\ref{Scwb} the
uniform covering condition of the previous section seems to be
awkward to verify directly, if true. It is therefore convenient to
proceed differently.  In what follows we will assume that $M$ is
an open domain in $\R^n$, this can again be gotten rid of by using
partitions of unity and passing to local charts.  As in the
preceding section, we assume that $\phi $ and $\varphi $ satisfy
\eq{threecB}-\eq{scalprop2} and that \eq{scalprop0} holds.

\begin{lem}\label{equivnormbis} The following
norms on  $C^{k,\alpha}_{\phi ,\varphi }(M)$ are equivalent:
$$
\begin{array}{lll}
\|u\|_{C^{k,\alpha}_{\phi ,\varphi }(M)}&\sim&\sup_{p\in M
}\|u\|_{C^{k,\alpha}_{\phi ,\varphi }(B_p)}\\
&\sim&\sup_{p\in M }\|u\|_{C^{k,\alpha}_{\phi(p) ,\varphi(p)
}(B_p)}.
\end{array}
$$
\qed
\end{lem}

For $u\in C^{k,\alpha}_{\phi ,\varphi }(M)$ and for
$\epsilon\in(0,1/2)$, we define the smooth function on $M$ (here
$\theta$ is as in the preceding section):
$$
\begin{array}{lll}
\widetilde{u}_\varepsilon(x)&=&\displaystyle{\int_{\R^n}
\frac{1}{\varepsilon^n\phi(x)^n}\theta\left(\frac{x-y}{\varepsilon\phi(x)}\right)u(y)d^ny}\\
&=&\displaystyle{\int_{y\in B(x,\varepsilon\phi(x))}
\frac{1}{\varepsilon^n\phi(x)^n}\theta\left(\frac{x-y}{\varepsilon\phi(x)}\right)u(y)d^ny}\\
&=&\displaystyle{\int_{z\in B(0,1)}
\theta(z)u(x-\varepsilon\phi(x)z)d^nz.}
\end{array}
$$
\begin{prop}\label{regulconvbis}
Let $u\in C^{k,\alpha}_{\phi ,\varphi }(M)$. For all
$\varepsilon\in(0,1/2)$ and all $m\in\N$ we have
$\widetilde{u}_\varepsilon\in C^{m}_{\phi ,\varphi }(M)$. Further,
$\widetilde{u}_\varepsilon$  converges to $u$ in
$C^{k,\alpha}_{\phi ,\varphi }(M)$ as $\varepsilon$ goes to zero.
\end{prop}

\proof We first show that $\widetilde{u}_\varepsilon\in
C^{k,\alpha}_{\phi ,\varphi }(M)$ with norm bounded independently
of $\varepsilon$. We have
$$
|\varphi(x)\widetilde{u}_\varepsilon(x)|\leq \sup_{y \in
B^n(x,\varepsilon\phi(x))} |\varphi(x)u(y)|\leq
C\|u\|_{C^{k,\alpha}_{\phi ,\varphi }(M)},$$ where the last
inequality comes from Lemma~\ref{equivnormbis}. For the first
derivatives, we have
$$
\partial_i\widetilde{u}_\varepsilon(x)=\displaystyle{\int_{z\in
B^n(0,1)}
\theta(z)\partial_ju(x-\varepsilon\phi(x)z)(\delta^j_i-\varepsilon\partial_i\phi(x)
z^j) d^nz.}$$ So from equation \eq{lcond} and
Lemma~\ref{equivnormbis} we have
$$
|\varphi(x)\phi(x)\partial_i\widetilde{u}_\varepsilon(x)|\leq
C\sup_{y \in B^n(x,\varepsilon\phi(x))} |\varphi(x)\phi(x)\partial
u(y)|\leq C'\|u\|_{C^{k,\alpha}_{\phi ,\varphi }(M)}.$$ It should
be  clear that similar inequalities are true for the $k^{th}$
derivatives and for the H\"older quotients, leading to
$$
\|\widetilde{u}_\varepsilon\|_{C^{k,\alpha}_{\phi ,\varphi
}(M)}\leq C\|u\|_{C^{k,\alpha}_{\phi ,\varphi }(M)}.$$ We leave it
as an exercise to the reader to    show in a manner similar to
that in the preceding section that $\widetilde{u}_\varepsilon$ is
in fact in $C^{k+l}_{\phi ,\varphi }(M)$ for all $l\in\N$.

Writing
$$
\widetilde{u}_\varepsilon(x)-u(x)=\displaystyle{\int_{z\in
B^n(0,1)} \theta(z)[u(x-\varepsilon\phi(x)z)-u(x)]d^nz,}
$$
one similarly shows that
$$
\|\widetilde{u}_\varepsilon-u\|_{C^{k,\alpha}_{\phi ,\varphi
}(M)}\leq C\|u\|_{C^{k,\alpha}_{\phi ,\varphi
}(M)}\varepsilon^\alpha\;,$$ so that $\widetilde{u}_\varepsilon$
converges to $u$ in $C^{k,\alpha}_{\phi ,\varphi }(M)$ when
$\varepsilon $ goes to zero, as required. \qed

\noindent {\sc Acknowledgements:} PTC acknowledges useful
discussions with L.~Andersson and R.~Bartnik.

\bibliographystyle{amsplain}
\bibliography{
../../references/newbiblio,%
../../references/reffile,%
../../references/bibl,%
../../references/hip_bib,%
../../references/netbiblio,../../references/addon}
\end{document}